\documentclass[a4paper,12pt]{article}

\usepackage{graphicx}
\usepackage[dvipsnames]{xcolor}
\usepackage[normalem]{ulem}
\usepackage{tgtermes}
\usepackage{array}
\usepackage{dsfont}
\usepackage[toc,page]{appendix}
\usepackage{setspace}
\usepackage{algorithm}
\usepackage{algpseudocode}
\usepackage{caption}
\usepackage{cite}
\usepackage{listings}
\usepackage{tikz,pgfplots}
\usepackage{hyperref}% http://ctan.org/pkg/hyperref
\hypersetup{%
  colorlinks = true,
  linkcolor  = black
}
\usepackage[utf8]{inputenc}
\usepackage{amsmath,amssymb,amsthm,textcomp}
\usepackage{enumerate}
\usepackage{multicol}
\usepackage{tikz}
\usetikzlibrary{positioning}

\usepackage[letterpaper, left=1in, right=1in, top=1in, bottom=1in]{geometry}

\newcommand{\Modelname}{Generalized Multinomial Logit model }

\newcommand{\bD}{ \mathbf{D} }

\newcommand{\bu}{ \mathbf{u} }
\newcommand{\bv}{ \mathbf{v} }

\newcommand{\reals}{ \mathbb{R} }

\def \Halmos{\mbox{\quad$\square$}}

\newtheorem{theorem}{Theorem}[section]

\newtheorem{lemma}[theorem]{Lemma}

\theoremstyle{definition}

\title{A Generalized Markov Chain Model to Capture Dynamic Preferences and Choice Overload}

\author{Kumar Goutam \thanks{Industrial Engineering and Operations Research, Columbia University, New York, NY., kg2621@columbia.edu} \and Vineet Goyal \thanks{Industrial Engineering and Operations Research, Columbia University, New York, NY. vgoyal@ieor.columbia.edu} \and Agathe Soret\thanks{Industrial Engineering and Operations Research, Columbia University, New York, NY.  acs2298@columbia.edu}}

\date{}

\pgfplotsset{compat=1.16}
\begin{document}

\maketitle

\begin{abstract}
Assortment optimization is an important problem that arises in many industries such as retailing and online advertising where the goal is to find a subset of products from a universe of substitutable products which maximize seller's expected revenue. One of the key challenges in this problem is to model the customer substitution behavior. Many parametric random utility maximization (RUM) based choice models have been considered in the literature. However, in all these models, probability of purchase increases as we include more products to an assortment. This is not true in general and in many settings more choices hurt sales. This is commonly referred to as the {\em choice overload}. In this paper we attempt to address this limitation in RUM through a generalization of the Markov chain based choice model considered in \cite{blanchet2016markov}. As a special case, we show that our model reduces to a generalization of MNL with no-purchase attractions dependent on the assortment $S$ and strictly increasing with the size of assortment $S$. While we show that the assortment optimization under this model is NP-hard, we present fully polynomial-time approximation scheme (FPTAS) under reasonable assumptions.

\vspace{2mm}
\textit{Key words}: assortment optimization, choice overload, choice model, Markov chain
\end{abstract}

\newpage

\section{Introduction} \label{sec_intro}

Assortment optimization problems arise in many practical applications such as retailing or online advertising. In such settings, a seller or decision maker has to select a subset of products from a universe of substitutable products to offer to customers in order to maximize the expected revenue. The demand of any item, and therefore the expected revenue, depend on the substitution behavior of the customers. Hence it is important to choose the ``right choice model" which specifies the probability that a random customer decides to purchase a particular item offered in the set. The objectives are two-fold: first determine or learn how customers choose and substitute among products, and second develop algorithms to find an optimal assortment.

More specifically, suppose we are given a universe of $n$ substitutable products, $N = \{1, ..., n \}$ with prices $p_1, ..., p_n$. Let $\pi(i,S)$ denote the probability that a random buyer selects product $i$ when the set of offered products is $S \subseteq N$. This probability depends on the substitution behavior of customers and is referred to as the choice probability. The assortment optimization problem, where goal is to find the subset $S$ of products that maximizes expected revenue, can be formulated as
$$
\max_{S \subseteq N} \sum_{i \in S} \pi(i,S) \cdot p_i.
$$
Many parametric choice models based on random utility maximization (RUM) have been studied in the literature to capture customer substitution behavior and the probability $\pi(i,S)$. 
% In RUM, the (random) utility of product $j$ for any customer $\tilde u_j = u_j + \xi_j$, where $u_j$ is the deterministic utility depending on product attributes and $\xi_j$ is the random utility that captures the idiosyncratic customer choice. 
% For any $S$, a customer selects product $j$ if $\tilde u_j \ge \tilde u_i, \forall i \in S\cup\{0\}$. Therefore we have $\pi(j,S) = \mathds{P}\left( \tilde u_j \ge \tilde u_i, \forall i \in S\cup\{0\} \right)$ where $0$ is the no-purchase option, which implies $\pi(j,S)$ is specified by the choice of the deterministic utility $u_j$ and the distribution of $\xi_j$.
One of the most popular choice model is the Multinomial Logit (MNL) model. The MNL model was introduced independently by \cite{luce2005individual} and \cite{plackett1975analysis}, and later studied by \cite{mcfadden1978modeling}. 
% In this model, the random component of the utility is assumed to be i.i.d. according to the standard Gumbel distribution. 
The probability that a customer purchases the product $i$ from assortment $S$ is given by 
$$
\pi(i,S) = \; \frac{e^{u_i}}{\sum_{j \in S} e^{u_j} + e^{u_0}} \mathds{1}_{i \in S} =: \; \frac{v_i}{\sum_{j \in S} v_j + v_0} \mathds{1}_{i \in S},
$$
where $v_j = e^{u_j}$. \cite{talluri2004revenue} show that both the estimation and the assortment optimization under this model are tractable. 
% More precisely, the optimal assortment is nested by price order, i.e. it is composed of the top $k$ priced products for some $k \le n$. 
Several algorithms 
% including greedy, local search and linear programming based methods 
are known to solve the assortment optimization problem under the MNL model
(\cite{talluri2004revenue}, \cite{gallego2004managing}, \cite{davis2013assortment}, and \cite{jagabathula2013two}). 
However, this model suffers from simplifying assumptions, such as the IIA - Independence of Irrelevant Alternatives property (see \cite{ben1985discrete}), which limit its applicability.
% in many practical settings.

Therefore, more complex choice models have been developed for capturing a richer class of substitution behaviors including the Nested Logit model (\cite{williams1977formation}, \cite{davis2014assortment}) and the mixture of Multinomial Logit model (\cite{RST14}). 
We refer the reader to \cite{berbeglia2018comparative} for a comprehensive survey.
In addition, ranking-based choice models of demand have also been studied in the literature (see \cite{jagabathula2013two}, \cite{farias2013nonparametric}, \cite{desir2016assortment}, \cite{jagabathula2017partial}, \cite{van2017expectation}). 
% With the whole spectrum of choice models, finding the right model to capture customer preferences in a particular application is a challenging problem. This is especially true since we only observe sales and not the complete preferences of the customers. There is a fundamental tradeoff between model complexity and the complexity of the resulting parameter estimation and assortment optimization problems.
\cite{blanchet2016markov} 
% address the model selection problem and 
present a Markov chain based model where the substitutions are modeled via transitions in a Markov chain. 
% They show that the Markov chain model captures the Multinomial Logit model exactly and provides a good approximation for any random utility based model under some fairly general assumptions.
% Furthermore, they show that the assortment optimization can be solved efficiently; thereby, providing  good balance between predictive power and tractability. 
This model was first considered in \cite{zhang2005revenue} and several variants 
% of this model 
have since been considered (\cite{chung2016re} and \cite{nip2017assortment}).

However, all standard random utility based models and distribution over ranking models suffer from two serious limitations in practice, namely, $i)$ the models assume that customer preferences are static and exogenous to the set of products offered by the seller, and $ii)$ the total probability of buying any product always increases (not necessarily strictly) if the seller adds more products to the assortment. In many settings, these properties are not satisfied. In particular, the customers may form their preferences based on the offered set of products and also, the purchase probability might decrease when the seller adds more products to the assortment. The latter is referred to as the {\em choice overload} phenomenon and has been observed empirically in practice (see \cite{iyengar2000choice} and \cite{scheibehenne2010can}). None of the random utility models in their fundamental and standard form can capture this. However, at least in principle, some modifications of existing methods might be able to achieve that. For example, one could potentially estimate a \textit{different} logit or mixed logit model for \textit{each} assortment offered to consumers, which would allow the preference parameters to vary with assortment. This seems to be a flexible way of letting the estimated demand functions depend on the products offered and thus should be able to capture choice overload. But in practice, this approach is not viable if the number of observations per assortment is small or the number of products itself is large. Hence more parsimonious models are a better option. \cite{wang2017impact} propose a model that incorporates an explicit search cost for users and can capture choice overload in some settings. \cite{wang2016consumer} also consider a choice model with endogenous network effects that capture dynamic preferences in some settings, mainly, where the utility of product for a customer depends on the number of customers interested in that product.

% \subsection{Related Work}
% In the Nested Logit model (see \cite{williams1977formation}), the products are partitioned into nests. However, both estimation and assortment optimization problems become more challenging under this model. For instance, the assortment optimization problem under this model is NP-hard in general (\cite{davis2014assortment}). \cite{davis2014assortment} show that under specific assumptions this problem is polynomially solvable. In the mixture of Multinomial Logit model, we consider the population to be a mixture of several segments, each of which is given by a MNL. The probability of selecting product $i \in S$ when the set $S$ is offered is given by
% $$
% \pi(i,S) = \sum_{k = 1}^K \alpha_k \frac{v^k_i}{\sum_{j \in S}^{} v^k_j + v^k_0},  
% $$
% where $K$ is the number of segments, $\alpha_k$ for all $k\in [K]$ denotes the probability that a random customer belongs to segment $k$ (hence $\alpha_1 + \ldots + \alpha_K = 1$), and $v^k \in \mathds{Q}_+^n$ for all $k \in [K]$ denote the MNL parameters for segment $k$. However, even for a mixture of MNL model with two segments, \cite{RST14} show that assortment optimization problem is NP-hard. Moreover, \cite{desir2017cap} show that it is hard to approximate within a factor better than $\Omega(n^{1 - \epsilon})$ in general. So the mixture of MNL model is quite intractable.

\subsection{Our Contributions} \label{our_contri}

The main goal of this paper is to develop a model for substitutions that addresses the above mentioned limitations and lead to a more practical framework for choice modeling and assortment optimization. We propose a generalization of the Markov chain model introduced in \cite{blanchet2016markov}, where we consider a Markovian comparison based choice process instead of one that is only based on Markovian substitutions. 
% We decide to extend and generalize this choice model as we believe 
The Markov chain choice model presents a natural framework to address these limitations as the substitution behavior of the customers are modeled 
% due to the way the customers make the choice (by doing a random walk on the offered products). Hence this is also naturally amenable to capture the search cost as well which in turn can model the customer preferences
in an intuitive manner making it a natural candidate.

\vspace{2mm}
\noindent \textbf{Our Model.} We assume that we are given a universe of $n$ substitutable products indexed from $1$ to $n$: $N = \{1, ..., n\}$. In our Markov chain based model, the customer substitutions are modeled using a Markov chain over $(n+1)$ states, $N_+ = \{0, 1, ..., n\}$: there is one state for each of the $n$ substitutable products and a state $0$ for the no-purchase alternative. We further assume that the transition probability matrix for the underlying Markov chain is given by $((\rho_{ij}))_{i\in N, j\in N_+}$. 
A customer starts at any random state $i$ and then does a random walk according to the transition probability matrix $\rho$. 
% defined on the Markov chain. 
When the customer arrives at a state $j$ which corresponds to product $j$ which is in the offer set $S$, they either buy the product and stop the random walk with a probability $\mu(j,S)$ which depends on other products in the offer set, or continue doing the random walk with the remaining probability according to the transition probability matrix until they reach either the state 0 (the no-purchase option) or any other state corresponding to a product in the offer set. We call this model as Generalized Markov chain choice model.

A key outcome of introducing such a stopping probability function is that this model is no longer an RUM model and hence the purchase probability may decrease when we add more products to the offer set, which we demonstrate via examples. 

\vspace{2mm}
\noindent {\bf Addressing Limitations}. In particular, our model addresses the discussed limitations in the following way:
\begin{itemize}
    \item {\bf Dynamic Preferences:} In practice, the substitution behavior of the customers in our model can be different for different assortments. More specifically, the implied distribution over preferences depends on the assortment offered  and there may not be any single distribution over preferences or rankings that is consistent with choices for all assortments. Our model captures this and to the best of our knowledge, this is the first systematic approach to capture dynamic preferences.
    
    \item {\bf Choice Overload Phenomenon:} An important consequence of our model is capturing the choice overload phenomenon. In particular, the probability of purchase does not necessarily increase if the seller includes more products in the assortment. More specifically, consider assortments $S, T \subseteq \{1,\ldots,n\}$ such that $S \subsetneq T$. Then it is not necessarily true that $\pi(0,S) \geq \pi(0,T)$, where $\pi(0,S)$ denotes the probability of no purchase when the set of products offered is $S$. We present several examples illustrating this.
\end{itemize}

\vspace{2 mm}
\noindent {\bf Generalized Multinomial Logit Model} We consider the special case of the Generalized Markov chain model where the underlying Markov chain has rank one and name it as the {\em Generalized Multinomial Logit model}. \cite{blanchet2016markov} show that the Multinomial Logit model can be exactly captured by a Markov chain model where the transition probability matrix has rank one. In particular, the choice probability under this model has the following expression: $$\pi(i,S) = \; \frac{v_i} {v_0 e^{\alpha \sum_{j\in S_+} v_j} + \sum_{j\in S} v_j},$$ where $\alpha \geq 0$ is a scale parameter. We can see that the no purchase probability is not constant but depends on the assortment as $v_0 e^{\alpha \sum_{j \in S_+} v_j}$, which increases the utility of the no-purchase alternative when the offer set is enlarged with more products.
    
\vspace{2 mm}
\noindent {\bf Assortment Optimization Under Our Model} 
We study the assortment optimization problem under the Generalized Markov chain choice model. Through examples, we demonstrate that an optimal assortment in our model balances between too few and too many choices and favors cluster centers. We show that the assortment optimization is NP-hard in general by a reduction from the partition problem. On the positive side, we present a fully polynomial time approximation scheme (FPTAS) for the assortment optimization problem when the transition probability matrix has either rank one or a constant rank more than one. 

Our algorithm for the FPTAS is based on exploiting the structure of the choice probability expression and consequently the expected revenue function. In particular, we show that the choice probabilities for a given assortment can be calculated via a system of linear equations. While the optimization problem of revenue maximization is non-convex, we show that we can obtain a convex approximation of the objective function by guessing the values of a small number of linear functions for the optimal assortment. Our algorithm is a dynamic programming based algorithm that adapts ideas from the dynamic programming algorithm for the knapsack problem.

\vspace{2 mm}
\noindent {\bf Outline} The rest of the paper is organized as follows. In Section \ref{sec_model_notations}, we present the Generalized Markov Chain model and our notations. In Section \ref{sec_choice_prob}, we present choice probability computations and we also discuss some examples. In Section \ref{sec_gmnl}, we present the properties of the Generalized Multinomial Logit model, a special case of the previous model along with an algorithm for its parameter estimation. 
Section \ref{sec_gmnl_fptas} discusses the assortment optimization under the Generalized MNL model and shows it is NP-hard. We also present a fully polynomial-time approximation scheme (FPTAS) for the assortment optimization problem. In Section \ref{sec_gmnl2}, we present results on another case of the Generalized Markov chain model, where the initial transition matrix is of low rank, and show that this generalizes the mixture of MNLs model. These results allow us to present an FPTAS for this model as well. Finally, we include some numerical results on real-life data in Section \ref{sec_numerics} and conclude in Section \ref{sec_concl}.

\section{Generalized Markov Chain Model} \label{sec_model_notations}

In this section we present the Generalized Markov Chain Model in details.

\subsection{Customer Substitution Behavior in the Markov Chain Model}

We model the customer substitution behavior using transitions on a Markov chain on $(n+1)$ states, where there is a state corresponding to each product and a state 0 for the no-purchase alternative.

We first describe the Markov chain choice model introduced in  \cite{blanchet2016markov}. 
Let $S \subseteq N$ be a subset of offered products, let $S_+ := S \cup \{0\}$. The model is specified by the parameters $\lambda_i$, $i\in [n]$ and $\rho_{ij}$, for all $i \in [n]$ and $j \in \{0, ..., n\}$. 

\begin{itemize}
    \item $\lambda_i$ is the arrival probability at state $i$: a customer starts at product $i$ with probability  $\lambda_i$,
    % enters the graph $\mathcal{G}$ with an arrival probability $\lambda_i$ at the vertex $i$,
    \item $\rho_{ij}$ denotes the transition probability from state $i$ to state $j$ if product $i$ is unavailable.
\end{itemize}
In the model introduced in \cite{blanchet2016markov}, when $S$ is offered, all the states corresponding to $S$ are absorbing. If the random walk of any customer reaches state $i \in S_+$, then they select product $i$ with probability one regardless of what else is being offered. 
% However, the assumption that the customer will buy product $i \in S$ with probability one if they reach it, implies that the model suffers from a certain limitation: indeed in this model, the customer preferences do not depend on the offered set. Therefore, if the seller offers more products, the customer will more likely reach a state in the offered set as now there are more absorbing states, thus decreasing the probability of reaching the no-purchase alternative. But as we pointed out above, this is not true in practice. In practice, when there are too many options, it is more difficult to make a decision, and therefore it is more probable to choose the no-purchase alternative. This is why we consider a Markovian model that captures customers' preferences.
This is again an RUM model and hence suffers from the limitations of choice overload and dynamic preferences as discussed previously.

\subsection{Substitution Behavior in Our Model}

In our model, we use the Markovian framework as above to model substitution behavior. However, we introduce a stopping probability function $\mu(i,S)$. 

\vspace{3 mm}
\noindent
{\bf Stopping probability function.} For any $i \in S$, $\mu(i,S)$ denotes the probability that a customer selects product $i$, given that they are currently already in state $i$ of the Markov chain. In the model considered by  \cite{blanchet2016markov}, this probability is equal to $1$. In this paper, we aim to capture the following fundamental component of customer choice, namely, that customer preferences and eventual selection depend on comparisons among the offered products. To capture this behavior, we model $\mu(i,S)$ as a decreasing function of $\sum_{j\in S} \rho_{ij}$ and consider the following formulation for $\mu(i,S)$
$$
\mu(i,S) = \exp \left( - \alpha \sum_{j \in N_+} \rho_{ij}x_j\right), \quad \forall i \in S,
$$
where $x_j = 1$ if $j\in S_+$ and 0 otherwise. Also, we have $\mu(i,S) = 0 \ \forall i \notin S$. Note that if a large number of products similar to $i$ (i.e. with large $\rho_{ij}$) are offered in the assortment, then the stopping probability is small. This reflects the scenario that it is difficult to select a product if a large number of similar options are available. Similarly, if we include more products in the assortment, $\mu(i,S)$ decreases. This models the fact that customers need more comparisons and time (transitions) to select the best product. This can be interpreted as the higher search cost for finding the best products if many similar items are offered.

We refer to this model as the {\em Generalized Markov Chain Model}. Here $\alpha$ is a scale parameter that amplifies the comparison effect. In our proposed model, we have $\alpha \ge 0$. A large value of $\alpha$ implies a very picky 
% and risk-averse 
customer. We would like to note that our model generalizes the model introduced in \cite{blanchet2016markov}. In particular, we recover the model in \cite{blanchet2016markov} by assuming $\alpha=0$.

Note that the choice of exponential function is arbitrary. We use this as it is a parsimonious choice to model transition probability. 
% But any other function that is decreasing in $\sum_{j \in N_+} \rho_{ij} x_j$ with range in $[0,1]$ could have worked as well. 

To model the eventual choice of product $i$ by the customer, we add a vertex $i'$, which is an absorbing state. A directed edge joins the vertex $i$ to the vertex $i'$ with weight $\mu(i,S)$, which represents the probability of buying the product $i$ when the customer is at the vertex $i$. This probability is equal to $0$ if and if only if the product is not offered, i.e. $i \notin S$.  We denote $N'_+$ to be the set of absorbing states $\{i'\}_{i \in [n]} \cup \{0\}$. After a certain time $t$, for $t$ large enough, the customer is either in a certain state $i'$, with $i \in S$, or in the no-purchase state $0$.

\vspace{3 mm}
\noindent
{\bf Modified transition probabilities.} Since the sum of the probabilities of getting out of $i$ has to be equal to $1$, we  change $\rho_{ij}$ to $\bar \rho_{ij}$ defined as follows:
   $$ \bar \rho_{ij} = (1 - \mu(i,S))\rho_{ij}.$$

\noindent
\textbf{Customer substitution behavior on the new graph.} Let us summarize how a customer behaves on the new graph given a certain set of products $S \subseteq N$ to sell under the Generalized Markov Chain model:
\begin{itemize}
\item The customer arrives with probability $\lambda_i$ at the vertex $i$.
\item If the product $i$ is in $S$ then the customer, currently at the vertex $i$,
\begin{itemize}
  \item either selects it with probability $\mu(i,S)$, arrives at the vertex $i'$ and then stops,
  \item or goes to another vertex $j$ with probability $\bar \rho_{ij}$.
\end{itemize}
  \item   If  $i \not \in S$, the customer cannot purchase $i$,
  so with probability $\bar \rho_{ij}$ they go to another vertex $j$.
  \item If $i = 0$, then the customer has decided not to purchase any product, and they stop. 
\end{itemize} 
We then proceed recursively.

\begin{figure}[!htb]
\centering
\begin{tikzpicture}
\draw[thick, draw=ForestGreen, fill=ForestGreen!5] (3, 0) circle (0.5cm) node[] {1};
\draw[thick, draw=ForestGreen, fill=ForestGreen!5] (3, 4) circle (0.5cm) node[] {2};
\draw[thick, draw=ForestGreen, fill=ForestGreen!5] (7, 4) circle (0.5cm) node[] {3};
\draw[thick, draw=ForestGreen, fill=ForestGreen!5] (7, 0) circle (0.5cm) node[] {4};
\draw[thick, draw=BrickRed, fill=BrickRed!5] (-0.5, -0.5) rectangle (0.5,0.5) node[pos=.5] {1'};
\draw[thick, draw=BrickRed, fill=BrickRed!5] (9.5, -0.5) rectangle (10.5,0.5) node[pos=.5] {4'};
\draw[thick, draw=BrickRed, fill=BrickRed!5] (-0.5, 3.5) rectangle (0.5,4.5) node[pos=.5] {2'};
\draw[thick, draw=BrickRed, fill=BrickRed!5] (9.5, 3.5) rectangle (10.5,4.5) node[pos=.5] {3'};
\draw[thick, draw=BlueViolet, fill=BlueViolet!5] (4.5, 1.5) rectangle (5.5,2.5) node[pos=.5] {0};
\draw[->, ForestGreen] (3.6,0) -- (6.4,0) node[pos=.5, below=0.1] {$\rho_{14}$};
\draw[->, ForestGreen] (3.6,3.8) -- (6.4,3.8) node[pos=.5, below=0.1] {$\rho_{23}$};
\draw[<-, ForestGreen] (3.6,4.2) -- (6.4,4.2) node[pos=.5, above=0.1] {$(1 - \mu(3,S))\rho_{32}$};
\draw[->, ForestGreen] (3,0.6) -- (3,3.4) node[pos=.5, left=0.1] {$\rho_{12}$};
\draw[->, BlueViolet] (3.45,0.45) -- (4.4,1.4) node[pos=.5, left=0.1] {$\rho_{10}$};
\draw[->, BlueViolet] (3.45,3.55) -- (4.4,2.6) node[pos=.5, left=0.1] {$\rho_{20}$};
\draw[->, BlueViolet] (6.55,0.45) -- (5.6,1.4) node[pos=.5, right=0.1] {$(1 - \mu(4,S))\rho_{40}$};
\draw[->, BlueViolet] (6.55,3.55) -- (5.6,2.6) node[pos=.5, right=0.1] {$(1 - \mu(3,S))\rho_{30}$};
\draw[->, BrickRed] (2.4,0) -- (0.6, 0) node[pos=.5, below=0.1] {$0$};
\draw[->, BrickRed] (2.4,4) -- (0.6, 4) node[pos=.5, below=0.1] {$0$};
\draw[->, BrickRed] (7.6,0) -- (9.4, 0) node[pos=.5, below=0.1] {$\mu(4,S)$};
\draw[->, BrickRed] (7.6,4) -- (9.4, 4) node[pos=.5, below=0.1] {$\mu(3,S)$};
\draw[->, black] (7,5.5) -- (7,4.6) node[pos=.5, right=0.1] {$\lambda_3$};
\draw[->, black] (3,-1.5) -- (3, -0.6) node[pos=.5, left=0.1] {$\lambda_1$};
\draw[->, black] (3,5.5) -- (3, 4.6) node[pos=.5, left=0.1] {$\lambda_2$};
\draw[->, black] (7,-1.5) -- (7, -0.6) node[pos=.5, right=0.1] {$\lambda_4$};
\end{tikzpicture}
\caption{\label{fig:exgraph}Example of a 4-vertex graph with $S = \{3,4\}$}
\end{figure}
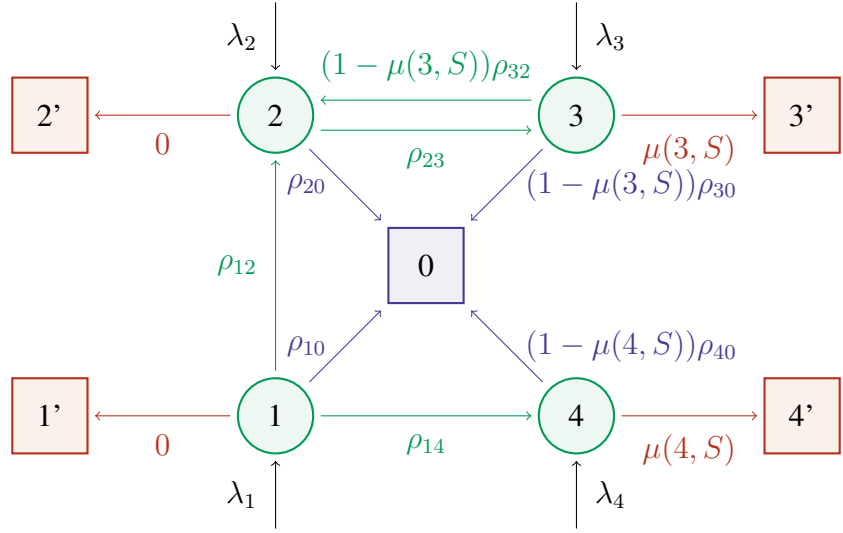

\vspace{2 mm}
\noindent
\textbf{Example} We consider the following 4-vertex graph (see Figure \ref{fig:exgraph}), where we have chosen to offer the subset $S = \{3,4\}$. Each product $i$ has a state $i$ as well as a state $i'$ where the latter is the absorbing state denoting that the customer buys product $i$.

We can see that since the offered set has products 3 and 4, the transition probabilities of going from node 3 or node 4 to any other node now gets modified, while other transition probabilities (i.e., from nodes 1 and 2 are unchanged). Firstly, nodes 3 and 4 are not absorbing anymore and hence the probability to go to nodes 3' and 4' which are now absorbing are $\mu(3,S)$ and $\mu(4,S)$ respectively. Secondly, this also changes the transition probabilities from nodes 3 and 4 to other nodes accordingly. 

% \subsection{Assortment Optimization Problem}
% Let $\pi(i, S)$ be the probability of buying the product $i$ when the subset $S$ is offered, and $p_i$ be the price of the   product  $i$. We finally assume that the seller has a fixed capacity of goods, therefore the cost price does not depend on our decision. The assortment optimization is the following 
% $$ \max_{S \subseteq N} \sum_{i \in S}  \pi(i, S) \times  p_i. $$
% Now the objective is to compute $\pi(i,S)$ under our model. 

\section{Computation of Choice Probabilities} \label{sec_choice_prob}

Given the parameters $\lambda_i$, $\bar \rho_{ij}$ and $\mu(i,S)$ for all $i \in N$ and $j \in N_+$, we can compute the choice probabilities for any $S \subseteq N$ in a very similar way as \cite{blanchet2016markov}. Our assumption is that a customer arrives at the state $i$ with probability $\lambda_i$, and continues to transition according to probabilities $\bar \rho_{ij}$ until they decide to buy a product $i$ with probability $\mu(i,S)$ when they are at the vertex $i$, or decide not to buy any product and end at the no-purchase vertex $0$. We therefore assume that any customer buys at most one product.

\subsection{Choice Probabilities}\label{choice_prob_form}

\noindent
Let $\rho(N, N)$ be the transition probability matrix from states $N$ to $N$. We recall that since the total probability of exiting a vertex $i$ is $1$, and since there is a probability $\mu(i,S)$ of buying the product represented by the vertex $i$, the transition probabilities are given by: $\bar \rho_{ij} = (1-\mu(i,S))\rho_{ij}$. Consequently,
$$
\rho(N,N) = Diag((1-\mu(i,S))) \times \rho,
$$
where $\rho$ is the initial transition probability matrix, $\rho = (\rho_{ij})_{i,j \in [n]}$, and $Diag((1-\mu(i,S)))$ is the diagonal matrix with $(1-\mu(i,S))$ on its diagonal. Also recall that $\mu(i,S) = 0 \ \forall i \notin S$.

After a certain time, every customer will be in an absorbing state. In order to compute $\pi(i, S)$, we have to know the probability that a customer arrives at the vertex $i$. 
For $i \in [n]$ we have:
$$ \pi(i, S) = \lim_{q \to \infty} \lambda^{T}(\mathcal{P}(S))^{q}e_{i}^{2n + 1}, $$
where $\mathcal{P}(S)$ is the transition probability matrix in the graph when the subset $S$ is offered and is of the form:
$$
\mathcal{P}(S) = \begin{bmatrix}
\rho(N'_+, N'_+) & \rho(N'_+, N) \\
\rho(N, N'_+) &
\rho(N, N)
\end{bmatrix} = \begin{bmatrix}
I_{n+1} & 0  \\
\Pi(S) & Diag((1-\mu(i,S)))\rho  \\
\end{bmatrix},
$$
and $\Pi(S)$ is the following matrix: 
$$
\Pi(S) = \rho(N, N'_+)
= 
\begin{bmatrix}
\mu(1,S) & 0 & ... & 0 & \bar \rho_{10} \\
0 & \mu(2,S) & ... & 0 & \bar \rho_{20} \\
: & & & : & : \\
0 & ... & 0 & \mu(n,S) & \bar \rho_{n0}
\end{bmatrix}
= \begin{bmatrix}
Diag(\mu(i,S)) & \bar \rho_0 \\
\end{bmatrix},
$$
 and $e_i^{2n + 1} \in \{0, 1\}^{2n + 1}$ has $0$ on each component except on its $i^{th}$ component. Since we have $i \in [n]$, $e_i^{2n + 1}$ always has its last $n + 1$ components equal to $0$. 
 $\rho(N'_+, N'_+) = I_{n+1}$ because all the states in $N'_+$ are absorbing, which also implies that $\rho(N'_+, N) = 0$.

For $q \in \mathbb{N}$, we have:
$$
\mathcal{P}(S)^q = \begin{bmatrix}
I_{n+1} & 0 \\
\sum_{k = 0}^{q}(Diag((1-\mu(i,S)))\rho)^k\Pi(S) & (Diag((1-\mu(i,S)))\rho)^q \\ 
\end{bmatrix}.
$$
Therefore, if we assume that the spectral radius of 
$\rho(N,N) = Diag((1-\mu(i,S)))\rho$ is strictly less than 1:
$$
\lim_{q \to \infty} \mathcal{P}(S)^q = \begin{bmatrix}
I_{n+1} & 0 \\
(I_{n} - Diag((1-\mu(i,S)))\rho)^{-1}\Pi(S) & 0 \\
\end{bmatrix},
$$
and  hence
$$
\pi(i,S) = \lambda^{T}(I_{n} - Diag((1-\mu(i,S)))\rho)^{-1}\Pi(S) e_i,
$$
where $e_i \in \{0,1\}^{n+1}$. Lastly, if we want the probability of no purchase, we can simply compute $\lambda^{T}(I_{n} - Diag((1-\mu(i,S)))\rho)^{-1}\Pi(S) e_0$, where $e_0 = (0, ..., 0, 1) \in \{0,1\}^{n+1}$.

\subsection{Examples} \label{sec_examples}

We now provide a couple of examples to show that our model is better at capturing the choice overload phenomenon than any other random utility based choice models. 

\vspace{2 mm}
\noindent {\bf Example 1 (Homogeneous Graph)}. We first consider a complete graph with $n$ vertices with homogeneous transition probabilities, $\rho_{ij} = \frac{1}{n+1}$ for all $i \in N$ and $j \in N_+$, and also homogeneous probabilities of arrival, $\lambda_i = \frac{1}{n+1}$ for all $i \in N_+$. We suppose that all products have same price $p$. The symmetry of this example implies that we only need to find the number $k$ of vertices to offer that would maximize revenue, and then randomly take a subset of $k$ elements. 
For any random utility based choice model presented in the first section, the optimal set to maximize revenue will be the entire set of products. Indeed, in all these models, the product with the highest price is always in the optimal assortment. Since all the products have the same price, they will all be in the set. Furthermore, the probability of no purchase always decreases when we add more products into the offered set.
However this is not true in our model, and the probability of no purchase depends on the parameter $\alpha$ chosen. We have chosen a universe of $n = 15$ substitutable products, and given the assumptions above, we compute the probability of purchase under our model when $k$ products are in the offered set, for all $k \in [n]$ and we obtain the graph in Figure \ref{fig:proba_nopurchase}:

\begin{figure}
\centering
\begin{tikzpicture}
    \begin{axis}[%
        axis x line=bottom,
        axis y line=left,
        xlabel=$\text{Number of products in the offered set}$,
        ylabel=$\text{Probability of no purchase}$,
        anchor= east]
        \addplot[mark=none,BrickRed,thick] table[x=x,y=a1] {probaNotbuy.dat};
        \addplot[mark=none,BlueViolet,thick] table[x=x,y=a2] {probaNotbuy.dat};
        \addplot[mark=none,ForestGreen,thick] table[x=x,y=a3] {probaNotbuy.dat};
        \legend{$\alpha = 1$,$\alpha = 2$, $\alpha = 10$}
    \end{axis}
\end{tikzpicture} 
\caption{Probability of no-purchase}
\label{fig:proba_nopurchase}
\end{figure}
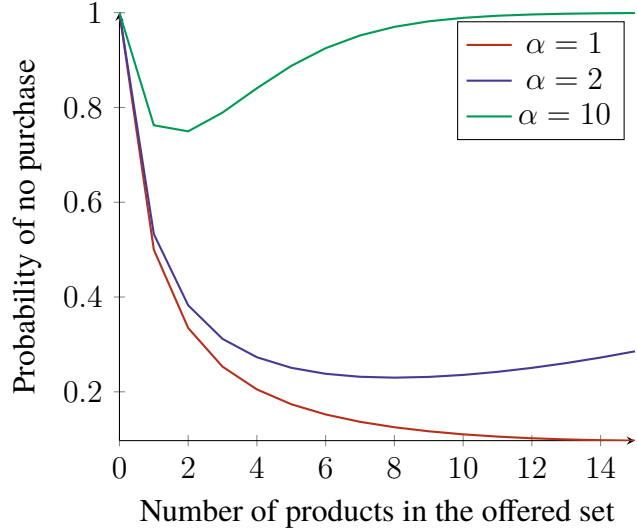

We see that, depending on the value of $\alpha$, the probability of no purchase may increase when we add more items into the offered assortment. Especially, a high $\alpha$ implies a sooner (in terms of the number of products) increase in the probability of no-purchase. 

We present another example, the star graph, which shows that our model will favour the cluster centers as items for the assortment set, unlike the Markov chain model. 

\vspace{2 mm}
\noindent
\textbf{Example 2 (Star Graph)}. We consider the following {\it star} graph with $n$ vertices. We suppose that vertex $1$ is linked to all other vertices in the graph, but other vertices are only linked to $1$ and to no purchase vertex $0$. We suppose that the transition probabilities are homogeneous. Therefore, they are given by:
$\rho_{1i} = \frac{1}{n} \ \forall i \in N_+ \backslash \{1\}$ and
$\rho_{i1} = \rho_{i0} = \frac{1}{2} \ \forall i \in N \backslash \{1\}$.
We suppose that the arrival probabilities are all equal $\lambda_i = \frac{1}{n} \ \forall i\in N$. Finally we suppose that all products, except $1$, have a price $P$, and product $1$ has a smaller price $p < P$. 

For any random utility based choice model considered in the literature, since the product with the highest price is always in the optimal offered set, then $\{2, ..., n\} \subseteq S^*$ where $S^*$ is the optimal set. And this is true, for any $n$, even for very large $n$. \\
However, our model considers that selling only the product $1$ will give a higher revenue, when $\alpha$ is large enough. For $\alpha \ge 9$, the optimal set will always be $\{1\}$. And this result is closer to the reality. Indeed, it seems more logical in practice for the seller to only offer the product which is similar to many other products, even if this product is slightly less expensive than the others.

\vspace{2 mm}
In order to get a generalization of the MNL model, we now suppose that the initial transition probability matrix, $\rho = (\rho_{ij})_{i,j \in [n]}$ is of rank one, and show that with such an assumption, the optimization problem is NP-hard.

\section{Generalized Multinomial Logit model} \label{sec_gmnl}

In this section we suppose that the transition probability matrix $\rho$ is of rank one. Given this assumption, we refer to our model as the Generalized Multinomial Logit model, which is a special case of the Generalized Markov chain model. We remind that $N$ represents all the vertices in the graph, $N_+$ is the union of all the vertices and the vertex $0$ which represents the no-purchase, and $\rho(N,N_+)$ is the transition probability matrix from $N$ to $N_+$. In this model, we suppose that there exists $\mathbf{v} = (v_i)_{i\in[n+1]}\in [0,1]^{n+1} =[v_0 \ \mathbf{v}_*^T]$, such that $\sum_{i = 0}^n v_i = 1$ and:
$$
\rho(N, N_+) = 
\begin{bmatrix}
1 - \mu(1,S)  &  & 0 \\
 &   \ddots &  \\
0 &   & 1 - \mu(n,S) \\
\end{bmatrix}
\times 
\begin{bmatrix}
v_0 & v_1 & ... & v_n \\
\vdots & \vdots & & \vdots \\
v_0 & v_1 & ... & v_n \\
\end{bmatrix}.
$$
Since there is a probability $\mu(i,S)$ that the customer goes from vertex $i$ to vertex $i'$, the probability that the customer goes from vertex $i$ to vertex $j$, with $j \in N_+$ , has to be $(1-\mu(i,S))v_j$, therefore the probability of exiting from vertex $i$ is $1$.
We also suppose in this model that $\lambda = \mathbf{v}$, therefore the probability of arriving at a vertex $i$ is proportional to $v_i$.
Finally we suppose that the probability of buying the product $i$ while being at vertex $i$ is given by: 
$$\mu(i,S) := e^{ - \alpha \times \sum_{j \in S_+} v_j }.$$

Additionally, we have $\mu(i,S) = 0 \ \forall i \notin S$. With the notations given in Section \ref{choice_prob_form}:
$$
\mathcal{P}(S) = \begin{bmatrix}
\rho(N'_+, N'_+) & \rho(N'_+, N) \\
\rho(N, N'_+) & \rho(N,N) \\
\end{bmatrix} = \begin{bmatrix}
I_{n+1} & 0 \\
\Pi(S) & D(S)\rho_v \\
\end{bmatrix},
$$
where 
$$
\Pi(S) = e^{- \alpha \sum_{j \in S_+} v_j} \times \begin{bmatrix}
\mathds{1}_{1 \in S} &   & 0 & (e^{ \alpha \sum_{j \in S_+} v_j} - \mathds{1}_{1 \in S})v_0\\
 &  \ddots &  & \vdots \\
0  &  & \mathds{1}_{n \in S} & (e^{ \alpha \sum_{j \in S_+} v_j} - \mathds{1}_{n \in S})v_0\\
\end{bmatrix} , \text{ and}
$$
\begin{align*}
D(S)\rho_v &= \begin{bmatrix}
1 - \mu(1,S) &  & 0 \\
  &  \ddots &  \\
0  &  & 1 - \mu(n,S) \\
\end{bmatrix}
\times 
\begin{bmatrix}
 v_1 & ... & v_n \\
\vdots &  & \vdots \\
 v_1 & ... & v_n \\
\end{bmatrix} 
= \left( 1 - e^{-\alpha \sum_{j \in S_+} v_j}\right) \times 
\begin{bmatrix}
 v_1 & ... & v_n \\
\vdots &  & \vdots \\
 v_1 & ... & v_n \\
\end{bmatrix}.
\end{align*}

We summarize here the assumptions of the Generalized Multinomial Logit model:

\vspace{2 mm}
\noindent
\textbf{Generalized Multinomial Logit Model} In this model we make the following assumptions:
\begin{itemize}
\item the initial transition probability matrix within $N_+$ is of rank one, i.e. there exists $\mathbf{v} = (v_i)_{i \in N_+} = \left( \begin{array}{l}
v_0 \\
\mathbf{v}_*
\end{array}\right) \in [0,1]^{n+1}$ such that
$\begin{array}{ll}
\rho(N,N) = Diag((1 - \mu(i,S)) \mathbf{1}\mathbf{v_*}^T & \mbox{ and } \sum_{j \in N} v_j + v_0 = 1
\end{array},
$
\
\item given a subset $S \subseteq N$, for all $i \in N$ we have $\mu(i,S) = e^{- \alpha \sum_{j \in S_+} v_j}$, \
\item and for all $j \in N_+$, $\lambda_j = v_j$.
\end{itemize}
\noindent
%\rule{\linewidth}{.5pt}

As we show in Section \ref{choice_prob_form}, the assortment optimization problem under our model is given by:
$$
\max_{S \subset N} \text{ }  v^T(I_n - D(S)\rho_v)^{-1}\Pi(S) p.
$$
We now give an exact formulation of choice probability and see why it generalizes the MNL model.

\subsection{Choice Probability}

We can compute the probability of choosing a product $i$ in our model as follows.

\begin{lemma} \label{gmnl_choice_prob}
The probability of purchasing a product $i$ given a chosen subset $S \subseteq N$ under the Generalized Multinomial Logit model is given by:
$$
\pi(i,S) = \frac{v_i}{\sum_{k \in S} v_k  + v_0e^{\alpha \sum_{j \in S_+} v_j}} \mathds{1}_{i \in S}.
$$
\end{lemma}
% \proof{Proof}
The proof follows from Section \ref{choice_prob_form} applied to this particular case and is presented in detail in Appendix \ref{gmnl_choice_prob_pf}.
%  \hfill \Halmos \endproof

\vspace{2 mm}
\noindent {\bf Our model is a generalization of the MNL model}. We recall that under the MNL model, the probability of buying the product $i \in S$ when the set $S$ is offered is given by
$$
\pi^{MNL}(i,S) = \frac{v_i}{\sum_{j \in S} v_j + v_0} \mathds{1}_{i \in S}.
$$
Therefore, our model can be considered as a generalization of the MNL model where the no purchase probability is not constant as is the case in MNL, but depends on the assortment $S$ as $v_0 e^{\alpha \sum_{j \in S_+} v_j}$, which increases the utility of the no-purchase alternative as compared to the MNL model.

Suppose that, instead of choosing, $\mu(i,S) = e^{- \alpha \sum_{j \in S_+} v_j}$, we had chosen a different function, say, $\mu(i,S) = \frac{1}{\sum_{j \in S_+} v_j}$, which would also convey the idea that $\mu(i,S)$ is a decreasing function of $\sum_{j \in S_+} v_j$. Then, the probability of purchasing the product $i$ given a set $S$ of offered products would have been:
$$
\bar \pi(i,S) = \frac{v_i}{\sum_{j \in S} v_j + v_0(\sum_{j \in S_+} v_j)} = \frac{v_i}{(1 + v_0) \sum_{j \in S} v_j + v_0^2},
$$
which in a ratio scale, is exactly the choice probability of MNL model. Therefore, such a function would have given a nesting by price order and therefore an optimization problem solvable in polynomial time, just as MNL. However, this model would not have given sufficient weights to the $v_j$'s for $j \in S$ on the no-purchase option, which is what we want to capture, namely the choice overload phenomenon. This is why we want to emphasize the importance of choice of the function $\mu(i,S)$ in our model, and why the choice of $e^{-\alpha \sum_{j \in S_+} v_j}$ meets the requirements of our model.

\subsection{Example}\label{examples}

Let us revisit the example of a homogeneous Markov chain from Section \ref{sec_examples} in this context.

\vspace{2 mm}
\noindent
\textbf{Example (Homogeneous Graph)}. We recall the assumptions in this model. Consider the case of the complete graph with $n$ vertices with homogeneous transition probabilities, $\rho_{ij} = \frac{1}{n+1}$ for all $i \in N$ and $j \in N_+$, and homogeneous probabilities of arrival, $\lambda_i = \frac{1}{n+1}$ for all $i \in N_+$. We suppose that all the products have the same price $p$. \\
For any random utility based choice model presented in the first section, the optimal set to maximize our revenue will be the entire set of products, as we explained before. This example is a particular case where the initial transition probability is of rank one. Therefore under the Generalized MNL model, the assortment optimization problem for the homogeneous graph is:
$$
\max_{S \subseteq N} \frac{\frac{|S|}{n+1}p }{\frac{|S|}{n+1} + \frac{1}{n+1} e^{\alpha \frac{|S|}{n+1}}} = \max_{k \in [n]} \frac{kp}{k + e^{\alpha \frac{k+1}{n+1}}}.
$$
A simple computation shows that the optimal number of products in the offered set is $k^* = \frac{n+1}{\alpha}$. Therefore, if $\alpha < \frac{n+1}{n}$ then the optimal assortment set will be the entire universe. However, if we take $\alpha$ large enough, then $\frac{n+1}{\alpha} \le n-1$ and there will be less products in the optimal offered set. This also highlights the meaning of $\alpha$: $\alpha$ amplifies the comparison effect. A large value of $\alpha$ implies risk-averse customer, and therefore a strategy where the seller should offer less products.

\subsection{Parameter Estimation for Generalized MNL Model} \label{est_gmnl}
Recall that the choice probabilities for the Generalized MNL model are given by:
\begin{align*}
    \pi(i,S)  = & \frac{v_i}{v_0e^{\alpha \sum_{j \in S_+} v_j} + \sum_{k \in S}
    v_k}, \quad \forall i \in S \\
    \pi(0,S)  = & \frac{v_0e^{\alpha \sum_{j \in S_+} v_j}}{v_0e^{\alpha \sum_{j
    \in S_+} v_j} + \sum_{k \in S} v_k},
\end{align*}
where $v_j=e^{\beta^\top x_j}$.

Given a choice dataset: $\bD=\{j_t,S_t\}_{t=1}^{T}$, where $S_t$ is the assortment set offered at time $t$ and $j_t$ is the choice made at time $t$ (which could be the outside option of no purchase), the log-likelihood can be formulated as:
% we have to estimate the parameters $\beta \in \bR^d$ and $\alpha>0$. To do a maximum likelihood estimation, we have to maximize the complete data log-likelihood:
$$ 
\ell(\bD,\beta,\alpha) = \sum_{t\notin \bD_0} \beta^\top x_{j_t} + \sum_{t\in \bD_0}
\big(\beta^\top x_0 + \alpha \sum_{j \in {S_t}_+} e^{\beta^\top x_j}\big) -
\sum_{t=1}^T \log\big(e^{\beta^\top x_0} e^{\alpha \sum_{j \in {S_t}_+} e^{\beta^\top
x_j}} + \sum_{k \in S_t} e^{\beta^\top x_k} \big),
$$
where $\bD_0$ is defined as the subset of $\bD$ where there was no purchase.

% \begin{lemma} \label{gmnl_est1}
% The log-likelihood function is not concave in $(\beta,\alpha)$ and hence the MLE
% problem is not a convex optimization problem.
% \end{lemma}

% In view of the above lemma, we prove the following statements and give an alternating
% projection maximization algorithm to get a local maximum.

While $\ell(\beta, \alpha)$ is not jointly concave in $(\beta,\alpha)$, we present an alternate algorithm based on searching for $\alpha$. In particular, we have the following lemma.

\begin{lemma} \label{gmnl_est2}
For a given value of $\alpha$, the maximization problem over $\beta$ can be
reformulated as a convex optimization problem. 
\end{lemma}
\proof{Proof}
The partial maximization problem over $\beta$ when $\alpha$ is known is the following:
$$
\max_{\beta} \sum_{t\notin \bD_0} \beta^\top x_{j_t} + \sum_{t\in \bD_0}
\big(\beta^\top x_0 + \alpha \sum_{j \in {S_t}_+} e^{\beta^\top x_j}\big)
- \sum_{t=1}^T \log\big(e^{\beta^\top x_0} e^{\alpha \sum_{j \in {S_t}_+}
e^{\beta^\top x_j}} + \sum_{k \in S_t} e^{\beta^\top x_k} \big).
$$
We introduce the following new variables:
$$
z_t = \beta^\top x_0 + \alpha \sum_{j \in {S_t}_+} e^{\beta^\top x_j}, \quad
t=1,\ldots,T
$$
Then we can re-write the above maximization as:
\begin{align}
\max_{\beta,z_t} & \sum_{t\notin \bD_0} \beta^\top x_{j_t} + \sum_{t\in \bD_0} z_t
- \sum_{t=1}^T \log\big(e^{z_t} + \sum_{k \in S_t} e^{\beta^\top x_k} \big)  \label{partial_beta} \\
\text{s.t. } & \beta^\top x_0 + \alpha \sum_{j \in {S_t}_+} e^{\beta^\top x_j} - z_t
= 0, \quad t=1,\ldots,T \notag
\end{align}
The objective function is now jointly concave in $(\beta,z_t)$ as it is a sum of linear functions
of $\beta$ and $z_t$ and the negative of log-sum-exp function. Also, the equality
constraints are convex functions in $(\beta,z_t)$. Hence we can solve this optimization problem in \eqref{partial_beta} efficiently. We also note that we are only introducing $T$ new variables and constraints.
 \hfill \Halmos \endproof

\begin{lemma} \label{gmnl_est3}
For a given value of $\beta$, the log-likelihood function is strictly concave in
$\alpha$ and hence it is unimodal. So the maximization problem over $\alpha$ can be solved.
\end{lemma}
\proof{Proof}
For a given value of $\beta$, the partial maximization problem over $\alpha$ is given by:
$$
\max_{\alpha}  \sum_{t\in \bD_0}  \alpha \sum_{j \in {S_t}_+} v_j
- \sum_{t=1}^T \log\big(v_0 e^{\alpha \sum_{j \in {S_t}_+} v_j} + \sum_{k \in S_t}
v_k \big)
$$
Defining $c_t:=\sum_{k \in S_t} v_k = \sum_{k \in S_t} e^{\beta^\top x_k}$, we can re-write this as
\begin{equation} \label{partial_alpha}
\max_{\alpha}  \sum_{t\in \bD_0} \alpha (v_0+c_t)
- \sum_{t=1}^T \log\big(v_0 e^{\alpha (v_0+
c_t)}+ c_t \big)    
\end{equation}

A simple derivative calculation shows the above function is strictly concave in
$\alpha$.
Hence the maximization problem in \eqref{partial_alpha} is also easy to solve. 
 \hfill \Halmos \endproof

\begin{figure}
 \centering
 \includegraphics[width=0.75\textwidth]{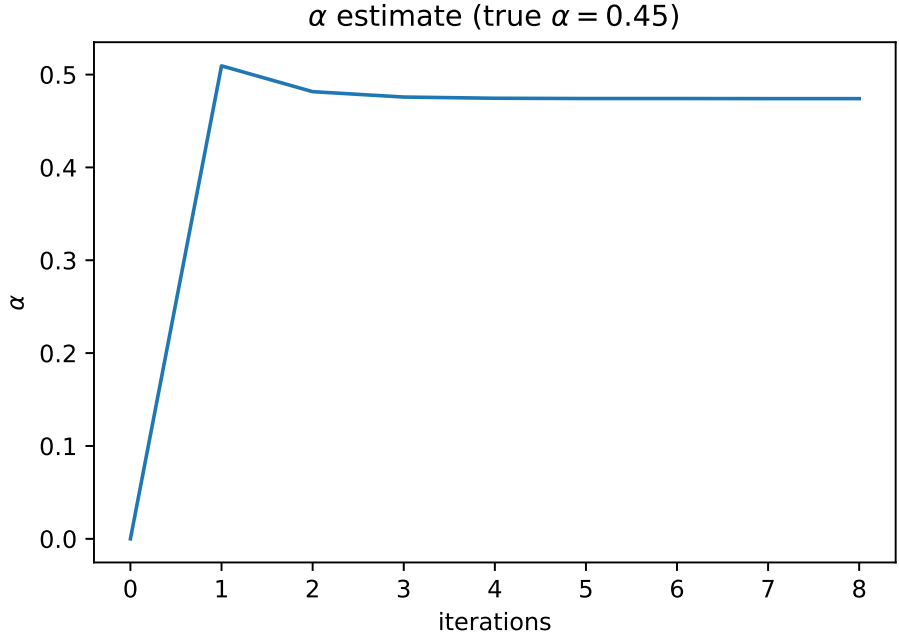}
 \caption{Estimate of $\alpha$ vs number of iterations for the estimation algorithm. The feature vector was 4 dimensional and we had 10 products, i.e., $d=4, n=10$. We can see convergence after a few iterations of the alternating projection algorithm.}
 \label{fig:alpha_estimate}
 \end{figure}

In accordance with the above results, an iterative algorithm for maximizing the
log-likelihood is to keep maximizing over $\alpha$ and $\beta$ alternatively until
convergence. This will lead to a local maximum. Since this type of alternative
maximization algorithm is dependent on the initial point, a good initial point could be
$\beta_{MLE}^{MNL}$ which is the maximum likelihood estimate for the MNL model (which
can be easily found as the MLE for MNL model is a convex optimization problem). The details are given in Algorithm \ref{gmnl_param_est}. We found Algorithm \ref{gmnl_param_est} to converge after a few iterations only, as evident from Figure \ref{fig:alpha_estimate}.

\begin{algorithm}
\caption{Parameter Estimation for \Modelname}\label{gmnl_param_est}
\begin{algorithmic}
\Procedure{ParamEstGenMNL}{$\bD$}
    \State Let $\beta_{MLE}^{MNL}$ be the maximum likelihood estimate from the given data $\bD$ for the MNL model \
    \State Set $\beta^{(0)} = \beta_{MLE}^{MNL}$
    \State Set $\alpha^{(0)} = 0$
    \For {$k=1,2,\ldots$} 
        \State Solve the optimization in \eqref{partial_alpha} with $\beta = \beta^{(k-1)}$ and set $\alpha^{(k)}$ to be the optimal solution \
        \State Solve the optimization problem in \eqref{partial_beta} with $\alpha = \alpha^{(k)}$ and set $\beta^{(k)}$ to be the optimal solution \
        \State Stop until convergence is achieved
    \EndFor
    \State Let $K$ be the index after convergence at the end of \textbf{for} loop
    \State Set $\beta_{MLE}^{GMNL} = \beta^{(K)}$
    \State Set $\alpha_{MLE}^{GMNL} = \alpha^{(K)}$
    \State \textbf{return} $\beta_{MLE}^{GMNL}$ and $\alpha_{MLE}^{GMNL}$ as the estimates
\EndProcedure
\end{algorithmic}
\end{algorithm}

\section{Assortment Optimization for the Generalized Multinomial Logit Model} \label{sec_gmnl_fptas}
In this section we consider the assortment optimization problem under the Generalized MNL model. Unlike the MNL model, even unconstrained assortment optimization under Generalized MNL model is NP-hard.
Under the \Modelname, using the expression of choice probability we derived in Lemma \ref{gmnl_choice_prob}, the assortment optimization problem can be written as
\begin{equation} \label{assort_gmnl}
\max_{S \subseteq N} R(S) := \; \max_{S \subseteq N} \; \frac{\sum_{i \in S} v_ip_i}{\sum_{i \in S} v_i + v_0e^{\alpha \sum_{i \in S_+} v_i}}.
\end{equation}

% In this section we prove that this problem is NP-hard, for a general $\alpha$. 

\subsection{NP-Hardness of the Assortment Optimization Problem}
In particular, we prove the following result. 
\begin{theorem} \label{gmnl_np_hard_gen}
The assortment optimization problem under the Generalized MNL model in \eqref{assort_gmnl} is NP-hard.
\end{theorem}
We use a reduction from the partition problem to prove this result and the details of the proof are presented in Appendix \ref{gmnl_np_hard_gen_pf}.

\vspace{2 mm} \noindent
Given that the assortment optimization problem in \eqref{assort_gmnl} is NP-hard, we can only hope to get an approximation. We present the best possible approximation in the form of a fully polynomial time approximation scheme (FPTAS).

Our algorithm for the FPTAS is based on the structure of the revenue function that depends on a single linear function of the assortment $S$, namely
$$V(S) = \sum_{j\in S}v_j.$$
In particular, for any assortment $S\subseteq [n]$, $R(S)$ is completely determined by $V(S)$. Hence if we can guess the value of $V(S^*)$ corresponding to an optimal assortment $S^*$, and find an assortment $S$ with $V(S) \approx V(S^*)$, we can use a dynamic programming based algorithm similar to the knapsack problem to construct such an approximately optimal assortment. This technique has been used in the past (see \cite{mittal2013general}). One of the most closely related works is \cite{desir2014near} which presents algorithms for constrained assortment optimization under many parametric models where the revenue function satisfies this linear structural property. 

\subsection{Our Algorithm for the FPTAS}

In this section, we now present a fully polynomial time approximation scheme (FPTAS) for the assortment optimization under the \Modelname \ discussed in \eqref{assort_gmnl}. For the FPTAS, we first consider different guesses for $V(S^*)$ in increasing powers of $(1+\epsilon)$. 

Let $v$ (resp. $V$) be the minimum (resp. maximum) value of the transition probabilities. We can assume that $v > 0$. For any given $\epsilon > 0$, we use the following set of guesses for $V(S^*)$:
$$
V_{\epsilon} = \{ v(1+\epsilon)^l, \; l=0,...,L \},
$$
where $L = O(\log(nV/v)/\epsilon)$. Hence, the number of guesses is polynomial in the number of products and $1/\epsilon$. 

Then for each guess $h \in V_{\epsilon}$, we consider discretized values of $v_j$ and try to construct an assortment $S$ such that:
$$h(1-\epsilon)\leq V(S) \leq h(1+\epsilon),$$ using a knapsack like dynamic programming. In particular, for  given guess $h \in V_{\epsilon}$, we try to find the best revenue possible with
$$
\sum_{j \in S} v_j \leq h,
$$
by a dynamic program. 

We consider the following discretized values of $v_j$ in multiples of $\epsilon h/n$:
$$ \forall j \in N \hspace{0.5cm} \bar v_j = \left\lceil \frac{v_j}{\epsilon h/n} \right\rceil.
$$
Let $I = \lceil n/ \epsilon\rceil + n$. For each $(i,k) \in [I] \times [n]$, let $R(i,k)$ be the maximum revenue of any subset $S \subseteq \{1, ..., k\}$ such that
$$
\sum_{j \in S} \bar v_j \leq i .
$$
We compute $R(i,k)$ using the following dynamic program
$$
R(i,1) = \left\{
\begin{array}{ll}
v_1p_1 & \mbox{ if } \bar v_1 \leq i \\
0 & \mbox{ if } i \ge 0 \\
-\infty & \mbox{ otherwise}
\end{array}
\right. 
$$
$$
R(i,k+1) = \max \{ v_{k+1}p_{k+1} + R(i - \bar v_{k+1}, k), R(i,k) \}.
$$
Let $S_h$ be the subset corresponding to $R(I,n)$, that is, the assortment $S_h$ that maximizes the sum $\sum_{j \in S} v_j p_j$ such that the inequality is verified for $i = I$. We then construct a set of candidate assortments $S_h$ for all guesses $h$, and return the best revenue that we get from all the candidates in the set. Algorithm \ref{FPTAS_gmnl} presents the details for the FPTAS.

\begin{algorithm}
\caption{FPTAS for the \Modelname }\label{FPTAS_gmnl}
\begin{algorithmic}
\Procedure{FPTASGenMNL}{$\epsilon, \mathbf{v}$}
    \For{$h \in V_{\epsilon}$} 
        \State Compute the discretized coefficients $\bar v_j = \left\lceil \frac{v_j}{\epsilon h/n} \right\rceil$ \
        \State Compute $R(i,k)$ for all $(i,k) \in [I] \times [n]$ using the dynamic program above \
        \State Let $S_h$ be the subset corresponding to $R(I,n)$
    \EndFor
    \State Let $\mathcal{C} = \cup_{h \in V_{\epsilon}} S_h$
    \State \textbf{return} the set $S^* \in \mathcal{C}$ that has the best revenue
\EndProcedure
\end{algorithmic}
\end{algorithm}

\begin{theorem} \label{gmnl_fptas}
Algorithm \ref{FPTAS_gmnl} returns a $(1- O(\epsilon))$-optimal solution to the assortment optimization problem in \eqref{assort_gmnl}. The running time is $O\left(\frac{n^2}{\epsilon^2}\log(nV/v)\right)$. 
\end{theorem}

We present the complete proof in Appendix \ref{gmnl_fptas_pf}.

%%%%%%%%%%%%%%%%%%%%%%%%%%%%

\iffalse%%%%%%%%%%%%%%%%%%%%%%%%%
We test the running time of our algorithm for different values of $n$ and $\epsilon$, as well as the relative error on the expected revenue.

\begin{center}
\begin{tabular}{|c|c|c|c|c|}
  \hline
   & \multicolumn{3}{c|}{Values of $\epsilon$} \\
  \cline{2-4}
   Values of $n$ & $10^{-2}$ & $5 \times 10^{-2}$ & $10^{-1}$\\
  \hline
  $5$  & 0.14 & & 0.01 \\
  $10$  & 4.2 & & 0.4 \\
  $15$ & 96.4 & 18.7 & 7.6 \\
  $20$ & 3656 & 621 & 375 \\
  \hline
\end{tabular} \\
\vspace{0.2cm}
{\bf Tab. 1} Running time (in seconds) for the FPTAS
\end{center}
Although the running times seem very high for $n \ge 20$, it is a great improvement, since the exact solution will be given in a dozen hours if we consider all the solutions. Furthermore, for $n \le 15$ and $\epsilon = 0.1$, the average relative error over $1000$ different observations is $0.4 \%$, which proves that this algorithm provides a very good estimation of the expected revenue.
\fi%%%%%%%%%%%%%%%%%%%%%%%%%%%%%%%%%%%

\section{Generalized Markov Chain Model with Low Rank} \label{sec_gmnl2}

% In the previous section, we supposed that the initial transition matrix is of rank one. 
In this section, we consider a general model, when the initial transition matrix is of low rank. In particular, we assume that the rank is some constant $K<n$ and the initial transition matrix $\rho$ is given by 
% \subsection{Model's Hypothesis}
% \noindent \textbf{Generalized Markov Chain Model with Low Rank Matrix} Let $K < n$, we suppose
% the initial transition matrix is of rank $K$. There exists $(\mathbf{u}_k)_{k \in [K]} \in ((0,1)^{n})^{K}$  and $(\mathbf{v}_{k})_{k \in [K]} \in ((0,1)^{n+1})^K$ such that for all $i \in [n]$, $\sum_{j = 0}^{n} \sum_{k = 1}^K u_{ik} v_{jk} = 1$, and the initial transition matrix in graph is given by
$$
\rho(N, N_+) = \left( \sum_{k \in [K]} u_{ik} v_{jk} \right)_{i \in [n], j \in N_+} = \sum_{k \in [K]} \mathbf{u}_k \mathbf{v}_k^T,
$$
where $\forall i \in [n], \; \sum_{j = 0}^{n} \sum_{k = 1}^K u_{ik} v_{jk} = 1.$
Given this initial transition probability matrix, we have that the probability of purchasing the product $i$ while being at vertex $i$ when $S$ is offered is
$$
\mu^{LR}(i,S) = e^{- \alpha  \left( \sum_{j \in S_+}  \sum_{k \in [K]} u_{ik} v_{jk} \right) } = e^{- \alpha  \left( \sum_{k \in [K]} u_{ik} V_k(S) \right) },
$$
where we define $$V_k(S) := \sum_{j \in S_+} v_{jk} \quad \forall k \in [K].$$

We also make the following assumptions:
\begin{itemize}
\item we suppose that $\forall j \in [n]$, $\forall k \in [K]$ $u_{jk} v_{jk} \le \frac{1}{n}$ (by this, we mean that the probability of staying at the state $j$ without buying the product $j$ cannot be too high);
\item we also suppose that $\alpha$ is not too large compared to $n$, more precisely, $\alpha \le \log n$.
\end{itemize}

First assumption is natural as it stipulates that a customer either buys the product or moves to another state. The second assumption is a technical assumption which makes sure that the spectral radius of the following matrix is bounded away from 1.

\begin{lemma}\label{lemma_rho_UV}
Let $S \subseteq N$, we define the following matrix $\in \mathcal{M}_K(\mathds{R})$:
$$M = UV(S) = \left( \sum_{j = 1}^{n} (1 - \mu^{LR}(j,S)) u_{jk} v_{jm} \right)_{k,m \in [K]}.$$
Then the spectral radius of $M$, 
$$\rho(M) \le 1 - \frac{1}{n^2}.$$
% (here $\rho(\cdot)$ denotes  a matrix).
\end{lemma}

% The fact that the spectral radius of $UV(S)$ is bounded away from $1$ will be of use in the next subsection and 
We present the proof in Appendix \ref{lemma_rho_UV_pf}.

\subsection{Assortment Optimization and FPTAS}

We present an FPTAS for the constant rank $K$ case with the running time of the algorithm being exponential in $K$. In particular, we first show that the expected revenue of an assortment $S$, $R(S)$ depends on $O(K)$ linear functions of $S$. Therefore if we can guess the values of these $O(K)$ linear functions for an optimal assortment $S^*$, and then find an assortment that approximately matches these values, we can compute the approximately optimal assortment.

Unlike other problems in literature (e.g., see \cite{mittal2013general}), the revenue function depends on a system of equations where the coefficients depend on the linear function values. Therefore to control the error in $R(S)$, we need to control the error in the estimates of the solution to the system of equations and not just the linear function values. This is one of the main challenges we address while constructing the algorithm for the FPTAS.

In particular, we choose the linear functions to guess more carefully, which allows us to give a theoretical bound on the error in solution of the system of equations.

We first compute the expected revenue of an assortment $S$ and give the following decomposition.

\begin{lemma} \label{gmnl2_exp_rev}
Under the Generalized Markov chain model with the rank of transition matrix being $K$, the expected revenue that we get from offering the assortment $S$ is
\begin{align*}
R^{LR}(S) = &\sum_{i \in [n]} \lambda_i (1 - \mu^{LR} (i,S)) \left(\sum_{j \in S} p_j \mu^{LR}(j,S) \bu_{i}^T [I - UV(S)]^{-1}\bv_{j} \right) + \sum_{i \in S} \lambda_i \mu^{LR}(i,S) p_i 
\end{align*}
% \begin{align*}
% R^{LR}(S) = &\sum_{i \in [n]} \lambda_i \sum_{k,k' = 1}^{K} u_{ik} (I - UV(S))^{-1}_{k'k}\left(\sum_{l = 1}^{n} v_l^{k'} \mu^{LR}(l,S) p_l\right) \\
% &+ \sum_{i \in S} \lambda_i \mu^{LR}(i,S) \left(p_i - \sum_{k,k' = 1}^{K} u_{ik} (I - UV(S))^{-1}_{k'k}\left(\sum_{l = 1}^{n} v_l^{k'} \mu^{LR}(l,S) p_l\right)\right),
% \end{align*}
% where $(I - UV(S))^{-1}_{k'k}$ is the coefficient of indices $k',k$ of the matrix $(I - UV(S))^{-1}$, 
where the matrix $UV(S) \in \mathcal{M}_{K}(\mathds{R})$ is defined by
$$
UV(S) = \left( \sum_{j = 1}^{n} (1 - \mu^{LR}(j,S)) u_{jk} v_{jm} \right)_{k,m \in [K]}.
$$
\end{lemma}

% \begin{align*}
% R^{LR}(S) = &\sum_{i \in [n]} \lambda_i \sum_{k,k' = 1}^{K} u_{ik} (I - UV(S))^{-1}_{k'k}\left(\sum_{\color{red}l \in S} v_l^{k'} \mu^{LR}(l,S) p_l\right) \\
% &+ \sum_{\color{red}i \in [n]} \lambda_i \mu^{LR}(i,S) \left(p_i - \sum_{k,k' = 1}^{K} u_{ik} (I - UV(S))^{-1}_{k'k}\left(\sum_{\color{red}l \in S} v_l^{k'} \mu^{LR}(l,S) p_l\right)\right) \\
% & {\color{red} - \sum_{i \in S^C} \lambda_i \mu^{LR}(i,S) p_i }
% \end{align*}

%  \begin{align*}
% R^{LR}(S) = &\sum_{\color{red}i \in S} \lambda_i \sum_{k,k' = 1}^{K} u_{ik} (I - UV(S))^{-1}_{k'k}\left(\sum_{\color{red}l \in S} v_l^{k'} \mu^{LR}(l,S) p_l\right) \\
% &+ \sum_{i \in S} \lambda_i \mu^{LR}(i,S) \left(p_i - \sum_{k,k' = 1}^{K} u_{ik} (I - UV(S))^{-1}_{k'k}\left(\sum_{\color{red}l \in S} v_l^{k'} \mu^{LR}(l,S) p_l\right)\right) \\
% & {\color{red} + \sum_{i \in S^C} \lambda_i (1-\mu^{LR}(i,S)) \left(\sum_{k,k' = 1}^{K} u_{ik} (I - UV(S))^{-1}_{k'k}\left(\sum_{l \in S} v_l^{k'} \mu^{LR}(l,S) p_l\right)\right) }
% \end{align*}

The proof builds from the general choice probability expression from Section \ref{choice_prob_form}. We present the details in Appendix \ref{gmnl2_exp_rev_pf}.
%  \hfill \Halmos \endproof

% In view of this theorem, if we define 
For any $i$ and $S$, let
\begin{equation}\label{f(i,S)}
    f(i,S) := \sum_{j \in S} p_j \mu^{LR}(j,S) \bu_{i}^T [I - UV(S)]^{-1}\bv_{j}.
    % \sum_{k,k' = 1}^{K} u_{ik} (I - UV(S))^{-1}_{k'k}\left(\sum_{l \in S} v_{lk'} \mu^{LR}(l,S) p_l\right),
\end{equation}
The assortment optimization problem under the Generalized Markov chain model can then be formulated as
\begin{equation}\label{assort_gmnl2}
   \max_{S \subseteq N} R^{LR}(S) := \; \max_{S \subseteq N} \; \sum_{i \in [n]} \lambda_i (1 - \mu^{LR} (i,S))f(i,S) + \sum_{i \in S} \lambda_i \mu^{LR}(i,S) p_i. 
\end{equation}

% \begin{align*}
% \max_{S \subseteq N} R^{LR}(S) := \max_{S \subseteq N} &\sum_{i = 1}^n \lambda_i \sum_{k,k' \in [K]} u_{ik} \left((I_K - UV(S))^{-1} \right)_{k'k} \left( \sum_{j = 1}^n v_{jk'}\mu^{LR}(j,S)p_j \right) \\
% &+ \sum_{i \in S} \lambda_i \mu^{LR}(i,S) \left(p_i -  \sum_{k,k' \in [K]} u_{ik} \left((I_K - UV(S))^{-1} \right)_{k'k} \left( \sum_{j = 1}^n v_{jk'}\mu^{LR}(j,S)p_j\right) \right)
% \end{align*}

% \subsection{FPTAS} \input{FPTAS2}

\subsection{FPTAS for the Generalized Markov Chain Model with Low Rank Matrix}

% We will present in this subsection an FPTAS for the  Generalized Markov chain model with low rank matrix. 
% We note here that the running time for our FPTAS in this low rank case is exponential in the fixed rank $K$, but this seems unavoidable in accordance with the hardness result.
% To prove that the FPTAS provides a $(1 - O(\epsilon))$-optimal solution to our assortment optimization problem, we will first need to prove the following lemma. 

We guess the following $K$ linear functions of $S$: 
$$V_k(S) = \sum_{j \in S_+} v_{jk} \quad \forall k \in [K].$$
We show the following result which stipulates that if our guesses are within $(1\pm \epsilon)$ of the optimal values, the error in solution of the system of equations is also within $(1\pm O(\epsilon))$.

\begin{lemma}\label{closeToUV}
Let $S \subseteq N$. Suppose that $\exists \ H 
% = (h_{kk'})_{k,k'\in [K]} 
\in \mathcal{M}_K(\mathds{R})$ 
% and $\exists \delta, \delta' > 0$ 
and $\hat{\bv} \in \mathds{R}$
such that 
% $\forall k,k' \in [K]$, 
% $$h_{kk'}(1 - \delta') \le UV(S)_{kk'}  \le h_{kk'}(1 + \delta).$$
$$(1 - O(\epsilon))H \le UV(S)  \le (1 + O(\epsilon))H \quad
\text{and} \quad (1 - O(\epsilon))\hat{\bv} \le \bv_j  \le (1 + O(\epsilon))\hat{\bv}.$$
Then we have that
% $$\left((I_K - H)^{-1}\right)_{kk'}( 1 - \delta' X(h)) \le \left((I_K - UV(S))^{-1} \right)_{kk'} \le \left((I_K - H)^{-1}\right)_{kk'}(1 + \delta X(H)),$$where $X$ is a function of the elements of $H$. 
$$
(1 - O(\epsilon))[I - H]^{-1}\hat{\bv} \le [I - UV(S)]^{-1}\bv_{j} \le (1 + O(\epsilon)) [I - H]^{-1}\hat{\bv}.
$$
\end{lemma}

The proof builds from Lemma \ref{lemma_rho_UV}. We present the details in Appendix \ref{closeToUV_pf}.
%  \hfill \Halmos \endproof

Now we are ready to present the FPTAS.
Let $v^k$ (resp. $V^k$) be the minimum (resp. maximum) value of $\{v_{ik}\}_{i \in N}$, for all $k \in [K]$. We can assume that $v^k > 0$. For any given $\epsilon > 0$, we use the following sets of guesses:
$$
W^k_{\epsilon} = \{ v^k(1+\epsilon)^t, t=0,...,T^k \}, \mbox{ for all } k \in [K],
$$
where $T^k = O(\log(nV^k/v^k)/\epsilon)$. A guess $h$ belongs in the set 
$$
W_{\epsilon} = W^1_{\epsilon} \times ... \times W^K_{\epsilon}.
$$
The number of guesses is polynomial in the input size and $1/\epsilon$. For  given guess $h = (h_1, ..., h_K) \in W_{\epsilon}$, we try to find the best revenue possible with
$$
h_k \le \sum_{j \in S_+} v_{jk} \leq h_k(1 + \epsilon), \mbox{ for all } k \in [K],
$$
using a dynamic program. In particular, consider the following discretized values of $v_{jk}$ in multiples of $\epsilon h_k/n$:
$$ \forall k \in [K], \hspace{0.4cm} \forall j \in N, \hspace{0.6cm} \bar v_{jk} = \left\lceil \frac{v_{jk}}{\epsilon h_k/n} \right\rceil.
$$
We denote by $\bar{\bv}_j$ the vector $\bar{\bv}_j:= (\bar v_{j1}, ..., \bar v_{jK})$. Let $L = \lceil n/\epsilon \rceil$, and $U = \lceil n/ \epsilon\rceil + n$. 
% We divided the revenue that we want to maximize in to two sums: one over all the elements in $N$, and the other one over the elements that are in $S$. 
We use a dynamic program to maximize the total expected revenue. 
% that we will denote as $R^{DP}$. 
For each $(l,u,m) \in [L]^K \times [U]^K \times [n]$, let $R^{DP}(l,u,m)$ be the maximum revenue of any subset $S \subseteq \{1, ..., m\}$ such that
$$
l_k \le \sum_{j \in S} \bar v_{jk} \leq u_k \hspace{0.5 cm} \forall k \in [K].
$$
For each guess $h$, let
$$
\mu_i(h) := e^{-\alpha \left(\sum_{k \in [K]} h_k u_{ik} \right)} \hspace{0.5 cm} \forall i \in N.
$$
% For each guess $h \in W_{\epsilon}$,
Therefore, $\mu(h) = (\mu_1(h), ..., \mu_n(h))$ is an estimate of the value of the $\mu^{LR}(i,S)$'s. Given this, we also use the following estimation $H$ of the matrix $UV(S)$ defined before:
$$
H(h) := \left(\sum_{i = 1}^n (1 - \mu_i(h))u_{ik} v_{ik'} \right)_{k,k' \in [K]}.
$$
Finally, let us also consider for each $i \in N$ and each $S$, the following estimate for $f(i,S)$  defined in equation (\ref{f(i,S)})
$$
f_i(h,S) = \sum_{j \in S} p_j \mu_j(h) \bu_{i}^T[I- H(h)]^{-1} \bv_{j}.
% f_i(h,S) = \sum_{k,k' = 1}^K u_{ik}(I- H(h))^{-1}_{k'k}\left( \sum_{j \in S} v_{jk'} \mu_j(h)p_j \right).
$$
For each guess $h$, we define the approximate revenue of the subset $S$ as
$$
R^{DP}(h,S) = \sum_{i\in S} \lambda_i \mu_i(h) p_i + \sum_{i=1}^n \lambda_i (1-\mu_i(h))f_i(h,S).
$$
We compute $R^{DP}(l,u,m)$ using the following dynamic program
$$
R^{DP}(l,u,1) = \left\{
\begin{array}{ll}
\lambda_{1} \mu_1(h)p_1 + \sum_{i=1}^n \lambda_i (1-\mu_i(h))
% \sum_{k,k' = 1}^K u_{ik}(I- H(h))^{-1}_{k'k} v_{1k'} \mu_1(h)p_1  
p_1 \mu_1(h) \bu_{i}^T[I- H(h)]^{-1} \bv_{1}
& \mbox{ if }  l \le \bar v_1 \leq u \\
0 & \mbox{ if } l \le 0 \mbox{ and } u \ge 0 \\
- \infty  & \mbox{ otherwise}
\end{array}
\right.
$$
\begin{align*}
R^{DP}(l,u,m) = \max \Big\{ & \lambda_{m} \mu_m(h)p_m + \sum_{i=1}^n \lambda_i (1-\mu_i(h)) 
p_m \mu_m(h) \bu_{i}^T[I- H(h)]^{-1} \bv_{m} \\
% \sum_{k,k' = 1}^K u_{ik}(I- H(h))^{-1}_{k'k} v_{mk'} \mu_m(h)p_m \\ 
& + R^{DP}(l - \bar v_{m}, u - \bar v_{m}, m-1), R^{DP}(l,u,m-1) \Big\}.
\end{align*}
Note that the number of states in the above dynamic program is  $O\left(\left(\frac{n}{\epsilon}\right)^{2K}n\right)$. Each step of the dynamic program requires a summation that can be done in $O(nK^2)$ time. This results in a total running time of $O\left(\left(\frac{n}{\epsilon}\right)^{2K}n^2 K^2\right)$.
% Let $S_h$ be the assortment corresponding to $R^{DP}(L,U,n)$. 
We construct a set of candidate assortments $S_h$ for all guesses $h$, and return the best revenue that we get from all the candidates in the set. 
Algorithm \ref{FPTAS_gmnl2} presents the details for the FPTAS.

\begin{algorithm}
\caption{FPTAS for the Generalized Markov chain model with Low rank matrix }\label{FPTAS_gmnl2}
\begin{algorithmic}
\Procedure{FPTASGenMixtMNL}{$\epsilon, \mathbf{u}_k, \mathbf{v}_k$}
    \For{$h \in W_{\epsilon}$} 
        \State Compute the discretized coefficients $\bar v_{jk} = \left\lceil \frac{v_{jk}}{\epsilon h_k/n} \right\rceil$ \
        \State Compute $R^{DP}(l,u,m)$ for all $(l,u,m) \in [L]^K \times [U]^K \times [n]$ using the dynamic program \
        \State Let $S_h$ be the subset corresponding to $R^{DP}(L, U, n)$
    \EndFor
    \State Let $\mathcal{C} = \cup_{h \in W_{\epsilon}} S_h$
    \State \textbf{return} the set $S^* \in \mathcal{C}$ that has the best revenue
\EndProcedure
\end{algorithmic}
\end{algorithm}
% Finally, we have to analyze the run time of this algorithm and show that it gives us a $(1- O(\epsilon))$ optimal solution which is proved next.
\begin{theorem} \label{fptas_gmnl2}
Algorithm \ref{FPTAS_gmnl2} returns a $(1- O(\epsilon))$-optimal solution to the assortment optimization problem in \eqref{assort_gmnl2}. The running time is $O\left( \frac{n^{2K + 2}}{\epsilon^{3K}}K^2\log(nV/v)^K\right)$. 
\end{theorem}

% \proof{Proof}
We present the proof in Appendix \ref{fptas_gmnl2_pf}.
%  \hfill \Halmos \endproof
Note that the running time is exponential in the fixed rank $K$ of the transition probability matrix.
% but this seems unavoidable in accordance with the hardness result.

\section{A Note on Feature Based Formulation for the Markov Chain Parameters}
Recall that the Generalized Markov chain model introduced in this paper has one more additional parameter ($\alpha$) in addition to the existing parameters of the Markov chain model. We first discuss the feature based formulation for the Markov chain model.
Consider the Markov chain choice model over $n$ substitutable products with features $x_1, \ldots, x_n \in {\mathbb R}^d$. The Markov chain choice model is given by first choice probabilities $\lambda_i$ for all $i \in [n]$ and transition probabilities, $\rho_{ij}$ for all $i\in [n]$ and $j \in \{0,1,\ldots,n\}$ where $0$ is the no-purchase or exit option. As such there are $O(n^2)$ parameters in the Markov chain model. This is significantly higher compared to say the widely used MNL model which only has $d$ parameters corresponding to the dimension of the attribute space. Recall the MNL choice model is given by attraction parameters, $v_i$ for all $i=1,\ldots,n$ where $v_i$ is a function of attributes $x_i$ of product $i$:
\[ v_i = e^{\beta^T x_i},\]
and the choice probability $\pi(i,S)$ for any assortment $S \subseteq [n]$ and $i \in S$ is given by:
\begin{equation}\label{eq:mnlform}
\pi(i,S) = \frac{ e^{\beta^T x_i} } { 1 + \sum_{j \in S} e^{ \beta^T x_j}}.
\end{equation}

Therefore, the MNL model is specified by parameters $\beta \in {\mathbb R}^d$. On the other hand, $O(n^2)$ parameters of the Markov chain model can be prohibitively large. This motivates us to consider a feature based formulation for the first choice probabilities, $\lambda_i$ and transition probabilities, $\rho_{ij}$ for all $i\in [n]$ and $j\in \{0,1,\ldots,n\}$. 

\vspace{2mm}
\noindent {\bf Feature based model for Transition Probabilities for MNL}.  We would like to first present a feature based formulation of the transition probabilities for the Markov chain model that represents a MNL model exactly. Consider a MNL model given by parameter $\beta \in {\mathbb R^d}$ where the choice probability for any assortment $S \subseteq [n]$ and $i\in S$ is given as in~\eqref{eq:mnlform}. Then consider the Markov chain model with the following feature based formulation of the parameters.
\begin{equation}\label{eq:MC-MNL-feature}
\begin{aligned}
\lambda_i & \propto \; e^{\beta^T x_i}, \; \forall i \in [n] \cup \{0\} \\
\rho_{ij} & \propto \; e^{\beta^T (x_j - x_i)}, \; \forall i \in [n], j \in [n]\cup \{0\}.
\end{aligned}
\end{equation}

Following the arguments in the proof of exact representation of a MNL model using a Markov chain model in~\cite{blanchet2016markov}, we can show the following.
\begin{theorem}\label{thm:MC-MNL-feature-exact}
Consider an MNL choice model with parameter $\beta\in {\mathbb R^d}$ over $n$ substitutable products with features $x_1, \ldots, x_n \in {\mathbb R^d}$. The Markov chain choice model corresponding to the feature based parameters defined in~\eqref{eq:MC-MNL-feature} is an exact representation of the MNL choice model.
\end{theorem}

\vspace{2mm}
\noindent {\bf Our Feature based model for Transition Probabilities}. The feature based formulation for transition probabilities in~\eqref{eq:MC-MNL-feature} is reasonable where the transition from product $i$ to product $j$ is proportional to the exponential of difference in utilities. However, since $e^{-\beta^T x_i}$ is common to all transitions out of product $i$, the transition probability, $\rho_{ij}$ just becomes proportion to $e^{\beta^T x_j}$. In other words, $\rho_{ij}$ only depends on $j$ and not on $i$ which leads to IIA and the formulation leading to representing the MNL model. Here we propose a slight variant of the feature based formulation that addresses this limitation and captures a significantly richer class of substitution behavior. In particular, we consider the following feature based form:

\begin{equation}\label{eq:MC-feature}
\begin{aligned}
\lambda_i & \propto \; e^{\beta_0^T x_i}, \; \forall i \in [n] \cup \{0\} \\
\rho_{ij}  & \propto \; e^{\beta_1^T (x_j - x_i)_+ + \beta_2^T (x_j-x_i)_-}, \; \forall i \in [n], j \in [n]\cup \{0\}.
\end{aligned}
\end{equation}

Here the parameters are $\beta_0, \beta_1, \beta_2 \in {\mathbb R}^d$. This is $O(d)$ parameters comparable to the MNL choice model. However, this formulation does not suffer from IIA and therefore, is able to model a significantly richer class of substitution behavior.

\section{Numerical Results} \label{sec_numerics}
In this section, we present numerical results on real data. We use the publicly available ``Related Article Recommendation Dataset" from \cite{beel2017rard} for performing this experiment.

\vspace{2 mm}
\noindent\textbf{Description.}
The dataset contains information on about 57.4 million recommendations that were displayed in the form of an ordered-list to the users of the digital library Sowiport. Information includes details such as which recommendation algorithms were used to order the list (one out of content-based filtering, stereotype, most popular and random) and also the time when those recommendations were requested, delivered and clicked.

From the digital library's point of view, the decision to be made is which articles, how many of them and in which \textit{order} should they be displayed when a request is received. The objective is to maximize the overall click-through rate (CTR), which is the ratio of clicked recommendations to those delivered. This dataset has also been used by \cite{beierle2017exploring} to study empirical evidence of the choice overload phenomenon. That study finds out that higher numbers of recommendations for a request lead to lower click-through rates.

\vspace{2 mm}
\noindent\textbf{Setup.}
Since the choice models assume that at most one product is selected from the offered assortment, we first filter out a few recommendations which had multiple clicks. After that, we do \emph{feature engineering} and build a few features which would be used to train the data on both MNL and Generalized MNL models. Then we fit both the models and estimate their respective parameters: $\beta \in \reals^d$ for the MNL model and $\beta \in \reals^d, \alpha > 0$ for Generalized MNL model. Since the parameter estimation for the MNL model is a convex optimization problem, we use the popular gradient descent method. We use the parameter estimation algorithm discussed in Section \ref{est_gmnl} for the Generalized MNL model, i.e., Algorithm \ref{gmnl_param_est}.

Once we have estimated the parameters for both the models, we use them to predict the click probabilities on a separate held-out dataset. After getting these click probabilities, we simply order the recommended articles for each request made, according to these values. Since the objective is to generate clicks on the recommended articles, we compare the predictions from both the models against actually clicked articles.

\begin{figure}
 \centering
 \includegraphics[width=0.8\textwidth]{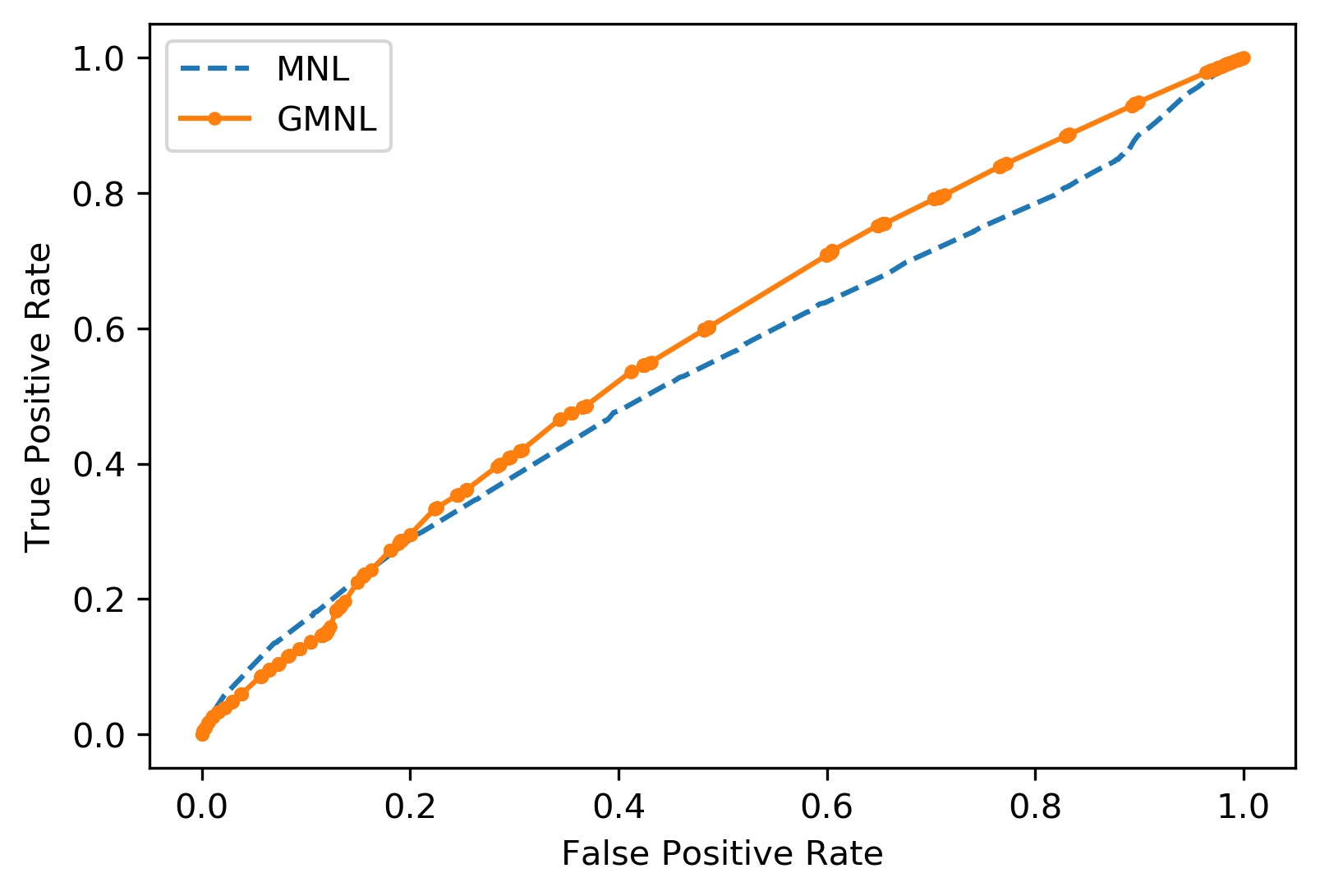}
 \caption{ROC Curves for MNL and Generalized MNL Models}
 \label{fig:roc_curve}
 \end{figure}

\vspace{2 mm}
\noindent\textbf{Results.}
The standard metric used in the literature when the objective is to maximize CTR is the area under the Receiver Operating Characteristic curve (the ROC AUC value, which lies between 0 and 1, with a higher value being preferable). We find out that Generalized MNL model improves the ROC AUC value over MNL model by 7\%. The plots of the ROC curves are shown in Figure \ref{fig:roc_curve}.

Another important observation is on the $\alpha$ estimated from the data. We get a high value for the estimate of $\alpha$ ($\alpha \approx 20$) which suggests that the choice overload phenomenon is prominent in such situations and hence Generalized MNL model would be able to capture them much better as compared to the MNL model.

\section{Conclusion} \label{sec_concl}
Our main contribution in this paper is to build upon the existing Markov chain based choice model presented by \cite{blanchet2016markov} and present a generalized model that addresses two significant limitations of existing random utility maximization and rank-based choice models in capturing dynamic preferences and the choice overload phenomenon. 

The Generalized Markov chain model attempts to capture both dynamic preferences and choice overload phenomenon by considering a modified choice or selection process, where a customer stops at a state corresponding to an offered product with some probability that depends on the set of offered products. This implicitly models the search cost in the selection process and therefore, captures both dynamic preferences and the choice overload phenomenon. Therefore, we present a novel framework to overcome the limitations in the existing choice models.

Considering the special cases when the transition matrix of the Markov chain is of rank one as well as when it has a low-rank, we show that the corresponding assortment optimization problem under these models is NP-hard. We also present a fully polynomial time approximation scheme (FPTAS) for both settings. 
The first model generalizes MNL model while the second model generalizes Mixture of MNLs model. 
We also show the effectiveness of Generalized MNL model on real data.

\section{Proofs of Various Lemmas and Theorems Appearing in the Paper}
\subsection{Proof of Lemma \ref{gmnl_choice_prob}} \label{gmnl_choice_prob_pf}
Let $S \subseteq N$, and $(I_n - D(S)\rho_v)^{-1} = (x_{ij})_{i,j \in N}$, our assortment optimization problem becomes:
\begin{align*}
    \max_{S \subset N}  \lambda^T(I_n - D(S)\rho_v)^{-1}\Pi(S)p & = \sum_{i \in S} \left( \sum_{k \in N} \lambda_k x_{ki} \right) e^{-\alpha \sum_{j \in S_+} v_j} p_i.
\end{align*}
Let $i \in S$ we want to compute
$$
\pi(i,S) = \left( \sum_{k \in N} \lambda_k x_{ki} \right) e^{-\alpha \sum_{j \in S_+} v_j} .
$$
We can show that:
$$
\begin{array}{ll}

\forall k,j \in N \hspace{0.5cm} &
\left\{ 
\begin{array}{ll}
  \frac{x_{kj}}{v_j(1 - \mu(k,S))} = \frac{1}{\sum_{s \in N} v_s\mu(s,S) + v_0} & \mbox{ if } k \ne j, \\
 \frac{x_{kk} - 1}{v_k(1 - \mu(k,S))} = \frac{1}{\sum_{s \in N} v_s\mu(s,S) + v_0} & \mbox{ otherwise. } 
\end{array}
\right.
\end{array}
$$

\noindent
Let $\pi_S = e^{-\alpha \sum_{j \in S_+} v_j}$ and $ x = \frac{1}{\sum_{k \in N} v_k\mu(k,S) + v_0} = \frac{1}{\pi_S \sum_{k \in S} v_k + v_0}$, then:

\begin{align*}
    \sum_{k \in N} \lambda_k x_{ki} & = \lambda_i x_{ii} + \sum_{k \ne i} \lambda_k x v_i (1 - \mu(k,S)) \\
    & = \lambda_i x_{ii} + x v_i \sum_{k \notin S} \lambda_k + x v_i (1 - \pi_S)\sum_{k \in S \backslash \{i\}} \lambda_k  \\
    & = \lambda_i x_ii + x v_i (1 - \lambda_i - \lambda_0) - x v_i \pi_S \sum_{k \in S \backslash \{i\}} \lambda_k \\
    & = \lambda_i(1 + (1-\pi_S)x v_i) + x v_i (1 - \lambda_i - \lambda_0) - x v_i \pi_S \sum_{k \in S \backslash \{i\}} \lambda_k \\
    & = \lambda_i + x v_i \left(1 - \lambda_0 - \pi_S \sum_{k \in S} \lambda_k \right)
\end{align*}

\noindent
Since we supposed that $\lambda_j = v_j$ for all $j \in N$, then the probability of buying the product $i$ becomes:
\begin{align*}
    \pi(i,S) & = \left( \sum_{k \in N} \lambda_k x_{ki} \right) e^{-\alpha \sum_{j \in S_+} v_j}  \\
    & = \pi_S v_i \left( 1 +  \frac{1 - v_0 - \pi_S \sum_{k \in S} v_k }{\pi_S \sum_{k \in S} v_k + v_0} \right) \\
    & = \pi_S v_i \left( \frac{1}{\pi_S \sum_{k \in S} v_k + v_0} \right) \\
   \pi(i,S) & = \frac{v_i}{\sum_{k \in S} v_k  + v_0 e^{\alpha \sum_{j \in S_+} v_j}}
\end{align*}
And if $i \notin S$, we have of course $\pi(i,S) = 0$ which finishes the proof.

\subsection{Proof of Theorem \ref{gmnl_np_hard_gen}} \label{gmnl_np_hard_gen_pf}
In this section, we present the details of the proof of the NP-hardness of the assortment optimization problem discussed in \eqref{assort_gmnl}.
We distinguish two different cases: $\alpha \le 1$ and $\alpha > 1$ based on a nice structural property of an optimal solution, when $\alpha \le 1$. 

\begin{lemma} \label{gmnl_alpha1}
In the \Modelname \ with $\alpha \le 1$, the product with the highest price is always in the optimal set, i.e. if $p_1 > p_2 \ge ... \ge p_n$, for all subset $S \subseteq N\backslash \{1\}$, we have
$$
R(S \cup \{1\}) \ge R(S) .
$$
\end{lemma}
\proof{Proof} If $N \backslash \{1\} = \emptyset$ then the result is trivial since $R(\emptyset) = 0$.
Now, suppose that there is at least one other product than $1$ in $N$. Let $S \subseteq N\backslash \{1\}$. We use the following notations:
$$
V(S) = \sum_{j \in S} v_j \hspace{0.5cm} \mbox{ and } \hspace{0.5cm} VP(S) = \sum_{j \in S} v_j p_j.
$$
Therefore we have
\begin{align*}
    R(S\cup \{1\}) - R(S) \ge 0 &\Leftrightarrow \frac{VP(S) + v_1 p_1}{V(S) + v_1 + v_0 e^{\alpha V(S_+)}e^{\alpha v_1}} - \frac{VP(S)}{V(S) + v_0e^{\alpha V(S_+)}} \ge 0\\
    &\Leftrightarrow v_1 (p_1 V(S) - VP(S)) + v_0e^{\alpha V(S_+)}(v_1p_1 - (e^{\alpha v_1} - 1)VP(S)) \ge 0
\end{align*}
Since $1$ has the highest price, $p_1 V(S) \ge VP(S)$ and
\begin{align*}
    (e^{\alpha v_1} - 1)VP(S)) &\le (e^{\alpha v_1} - 1) V(S) p_1 = (e^{\alpha v_1} - 1)(1 - v_1 - \beta(S))p_1,  
\end{align*}
where $ \beta(S) = \sum_{k \in N_+ \backslash \{\{1\}\cup S\}} v_k$. Moreover, since
$$(e^{\alpha v_1} - 1)(1 - v_1 - \beta(S)) - v_1 = e^{\alpha v_1}(1 - v_1 - \beta(S)) - 1 + \beta(S),$$ we want to prove that $g : x \mapsto e^{\alpha x}(1 - x - \beta(S))-1$ is a negative function on $(0,1)$. Indeed, it is a strictly decreasing function on $(0,1)$:
$$
g'(x) = e^{\alpha x}(\alpha - \alpha x - \alpha \beta(S) - 1) < 0 \Leftrightarrow 1 - \beta(S) -x < \frac{1}{\alpha}.
$$
And $1 - \beta(S) - x = V(S) \le 1 - v_0 < \frac{1}{\alpha}$ since we assumed $\alpha \le 1$.
Moreover, $g(0) = 0$, therefore $g$ is negative on $(0,1)$ and we have $R(S\cup \{1\}) - R(S) \ge 0$.
\hfill \Halmos \endproof

\vspace{2 mm}

Once we have this result, we now make a reduction from the partition problem. 
Consider the following instance of the partition problem: we are given $n$ integers $c_1, ..., c_n$ and the goal is to decide whether there is a subset $S \subseteq \{1, ...,n \}$ such that $\sum_{i \in S} c_i = \sum_{i \in \{1, ..., n\} \backslash S} c_i$.

\noindent
Let $T = \frac{1}{2} \sum_{i=1}^n c_i$, then $\sum_{i \in S} c_i = \sum_{i \in \{1, ..., n\} \backslash S} c_i$ if and only if $\sum_{i \in S} c_i = T$. We can suppose without loss of generality that $c_i > 0$ for all $i \in [n]$. We construct an instance of our problem as follows: 
$$
v_i = \left\{ \begin{array}{ll}
\frac{c_i}{2T + 1} & \mbox{ if } i \ge 1, \\
1 - \sum_{ i = 1}^n v_i = \frac{1}{2T+1} & \mbox{ if } i = 0. \\
\end{array}
\right.
$$
and let $c_0 := v_0 e^{\alpha v_0} > 0$, we define the prices as follows
$$
p_i = \left\{ \begin{array}{ll} 
\frac{1}{(2T+1)c_0} + \frac{e^{\frac{\alpha T}{2T+1}}-1}{T} + \frac{1}{c_i} &\mbox{ if } i = 1, \\
\frac{1}{(2T+1)c_0} + \frac{e^{\frac{\alpha T}{2T+1}}-1}{T} & \mbox{ otherwise. }
\end{array}
\right.
$$
Finally we set the target revenue as $K = \frac{\frac{T}{(2T+1)c_0}+e^{\frac{\alpha T}{2T+1}}}{T + (2T+1)c_0e^{\frac{\alpha T}{2T+1}}}$. \\
First, we can note that $1$ is necessarily in the optimal set. Indeed $1$ has the highest price in $N$, so the previous lemma implies that $1$ is necessarily in the optimal set (we can note that the choice of $1$ is random and we could have chosen any $i$ in $N$).

\noindent
In this case, our problem becomes
\begin{align*}
    \max_{S \subseteq \{1,...,n\}} R(S) &:= \max_{S\subseteq \{2,...,n\}} R(S \cup \{1\}) \\
    &= \max_{S \subseteq \{2,...,n\}} \frac{\sum_{i \in S} v_i p_i + v_1p_1}{\sum_{i \in S\cup \{1\}} v_i + c_0 e^{\alpha \sum_{i \in S\cup \{1\}} v_i}} \\
    &= \max_{S \subseteq \{1,...,n\}} \frac{ \left(\frac{1}{(2T+1)c_0} + \frac{e^{\frac{\alpha T}{2T+1}}-1}{T} \right) \sum_{i \in S} c_i + 1}{\sum_{i \in S} c_i + (2T + 1)c_0 e^{\frac{\alpha}{2T + 1} \sum_{i \in S} c_i}}\\
    &=: \max_{S \subseteq \{1,...,n\}} F\left(\sum_{i\in S} c_i\right) 
\end{align*}
where 
$$F : \begin{array}{ll} [0,2T] &\rightarrow \mathds{R}_+ \\ x &\mapsto \frac{h(x)}{x + (2T+1)c_0 e^{\frac{\alpha}{2T+1} x}} \end{array},$$
and  
$$h : x \mapsto \left(\frac{1}{(2T+1)c_0} + \frac{e^{\frac{\alpha T}{2T+1}}-1}{T} \right)x + 1.$$
$F$ is increasing at $x$ if and only if
\begin{align*}
    F'(x) \ge 0 &\Leftrightarrow \frac{h'(x)(x + (2T+1)c_0 e^{\frac{\alpha}{2T+1} x}) - h(x)(1 + c_0 \alpha e^{\frac{\alpha}{2T+1} x})}{(x + (2T+1)c_0 e^{\frac{\alpha}{2T+1} x})^2} \ge 0 \\
    &\Leftrightarrow h'(x) - \frac{1 + c_0 \alpha e^{\frac{\alpha}{2T+1} x}}{x + (2T+1)c_0 e^{\frac{\alpha}{2T+1} x}} h(x) \ge 0 \\
    &\Leftrightarrow \frac{h'(x)}{h(x)} \ge \frac{1 + c_0 \alpha e^{\frac{\alpha}{2T+1} x}}{x + (2T+1)c_0 e^{\frac{\alpha}{2T+1} x}} > 0  \hspace{0.3cm} \mbox{ since } h > 0 \mbox{ on } [0,1] \\
    &\Leftrightarrow \ln(h(x)) - \ln(h(0)) \ge \ln(x + (2T+1)c_0 e^{\frac{\alpha}{2T+1} x}) - \ln((2T+1)c_0) \\
    &\Leftrightarrow h(x) \ge \frac{1}{(2T+1)c_0} (x + (2T+1)c_0 e^{\frac{\alpha}{2T+1} x}) \hspace{0.5cm} \mbox{ since } h(0) = 1 \\
    &\Leftrightarrow  h(x) - g(x) \ge 0
\end{align*}
where $g : x \mapsto \frac{1}{(2T+1)c_0} (x + (2T+1)c_0 e^{\frac{\alpha}{2T+1} x})$. We note that $g$ is a strictly increasing function on $[0,2T]$ such that $h(0) = g(0) = 1$ and $h(2T) < g(2T)$. Indeed, 
\begin{align*}
h(2T) - g(2T) = 2 (e^{\frac{\alpha T}{2T+1}} - 1) + 1 - e^{\frac{\alpha \times 2T}{2T+1}} &= - (e^{\frac{\alpha T}{2T+1}} - 1)^2 < 0.
\end{align*}
Therefore, since $h$ is a line with a positive slope, there exists a unique $x^* \in (0,2T)$ such that for all $0 < x < x^*$, $h(x) - g(x) > 0$, $h(x^*) - g(x^*) = 0$ and for all $2T \ge x > x^*$, $h(x) - g(x) < 0$. But $h(T) =  g(T)$. So $x^* = T$. So this proves that $F$ is strictly increasing on $[0,T)$ then strictly decreasing on $(T, 2T]$. So $F$ has a unique maximum at $T$ on $(0,2T)$. Hence,
$$
    \max_{S \subseteq \{1,...,n\}} R(S) = \max_{S \subseteq \{1,...,n\}} F\left( \sum_{i \in S} c_i \right)  \le F(T) 
     = \frac{\frac{T}{(2T+1)c_0}+e^{\frac{\alpha T}{2T+1}}}{T + (2T+1)c_0e^{\frac{\alpha T}{2T+1}}} = K .
$$
So there exists an assortment $S \subseteq \{1,...,n\}$ whose expected revenue is at least $K$ if and only if the chain of inequalities hold as equalities. For this to happen we need to have $\sum_{i \in S'} c_i = T$ for some assortment $S' \subseteq \{1,...,n\}$. Therefore there exists an assortment $S \subseteq \{1,...,n\}$ whose expected revenue is at least $K$ if and only if there exists an assortment $S' \subseteq \{1,...,n\}$ that satisfies $\sum_{i \in S'} c_i = T$. 

\vspace{3 mm}

Although when $\alpha > 1$, we no longer have any nice structure in the optimal assortment and the product with the highest price \emph{may not} necessarily be in an optimal assortment set, we can still prove that the assortment optimization problem is NP-hard as long as $\alpha > 2$ (recall that we are interested in the high $\alpha$ case).

% \begin{theorem} \label{gmnl_np_hard2}
% The optimization problem under the Generalized MNL model is NP-hard for $\alpha > 2$.
% \end{theorem}

% \proof{Proof}
\vspace{3 mm}
To prove the NP-hardness in this case, we  once again make a reduction from the partition problem. Consider the following instance of the partition problem: we are given $n$ integers $c_1, ..., c_n$ and the goal is to decide whether there is a subset $S \subseteq \{1, ...,n \}$ such that $\sum_{i \in S} c_i = \sum_{i \in \{1, ..., n\} \backslash S} c_i$.

\noindent
Let $T = \frac{1}{2} \sum_{i=1}^n c_i$, then $\sum_{i \in S} c_i = \sum_{i \in \{1, ..., n\} \backslash S} c_i$ if and only if $\sum_{i \in S} c_i = T$. We can suppose without loss of generality that $c_i > 0$ for all $i \in [n]$. We construct an instance of our problem as follows: 
$$
v_i = \left\{ \begin{array}{ll}
\frac{c_i}{T \times \alpha} & \mbox{ if } i \ge 1, \\
1 - \sum_{ i = 1}^n v_i = 1 - \frac{2}{\alpha} & \mbox{ if } i = 0, \\
\end{array}
\right.
$$
and let $c_0 := v_0 e^{\alpha v_0} > 0$. We note that $v_0 > 0$ because we have supposed $\alpha > 2$. We define the prices as follows
$$
\forall i \in [n] \hspace{0.5cm} p_i = 1 .
$$
Finally we set the target revenue as $K = \frac{1}{1 + \alpha c_0 e}$. In this case, our problem becomes
\begin{align*}
    \max_{S \subseteq \{1,...,n\}} R(S) 
    &:= \max_{S \subseteq \{1,...,n\}} \frac{\sum_{i \in S} v_i p_i}{\sum_{i \in S} v_i + c_0 e^{\alpha \sum_{i \in S} v_i}} \\
    &= \max_{S \subseteq \{1,...,n\}} \frac{ \frac{1}{T \alpha }  \sum_{i \in S} c_i}{\frac{1}{T \alpha}\sum_{i \in S} c_i + c_0 e^{\frac{1}{T} \sum_{i \in S} c_i}} \\
    &=: \max_{S \subseteq \{1,...,n\}} F\left(\sum_{i\in S} c_i\right),
\end{align*}
where 
$$F : \begin{array}{ll} [0,2T] &\rightarrow \mathds{R}_+ \\ x &\mapsto \frac{x}{x + T \alpha c_0 e^{\frac{x}{T}}} \end{array}.$$
$F$ is increasing at $x$ if and only if
\begin{align*}
    F'(x) \ge 0 &\Leftrightarrow \frac{x + T \alpha c_0 e^{\frac{ x}{T}} - x(1 + \alpha c_0 e^{\frac{ x}{T}})}{(x + T \alpha c_0 e^{\frac{ x}{T}})^2} \ge 0 \\
    &\Leftrightarrow x \le T .
\end{align*}
Therefore $F$ is strictly increasing on $[0,T)$ then strictly decreasing on $(T, 2T]$. So $F$ has a unique maximum at $T$ on $(0,2T)$. Hence,
$$
    \max_{S \subseteq \{1,...,n\}} R(S) = \max_{S \subseteq \{1,...,n\}} F\left( \sum_{i \in S} c_i \right)  \le F(T) 
     = \frac{1}{1 + \alpha c_0 e} = K .
$$
So there exists an assortment $S \subseteq \{1,...,n\}$ whose expected revenue is at least $K$ if and only if the chain of inequalities hold as equalities. For this to happen we need to have $\sum_{i \in S'} c_i = T$ for some assortment $S' \subseteq \{1,...,n\}$. Therefore there exists an assortment $S \subseteq \{1,...,n\}$ whose expected revenue is at least $K$ if and only if there exists an assortment $S' \subseteq \{1,...,n\}$ that satisfies $\sum_{i \in S'} c_i = T$. 

%  \hfill \Halmos \endproof
\vspace{2mm}
 
We make a special note here about $1< \alpha \leq 2$. We point out that we are more interested in the ``picky customer" case, i.e., when $\alpha$ is large enough, because this is when we actually characterize choice overload (see the homogeneous graph example in Section \ref{examples}, or the numerical results in Section \ref{sec_numerics}). Hence this is an uninteresting case, although we do believe that assortment optimization problem is still NP-hard for this particular setting as well (and we have examples to verify this claim).

% \subsection{Proof of Lemma \ref{gmnl_alpha1}}\label{gmnl_alpha1_pf}

% \subsection{Proof of Theorem \ref{gmnl_np_hard1}}\label{gmnl_np_hard1_pf}

% \subsection{Proof of Theorem \ref{gmnl_np_hard2}}\label{gmnl_np_hard2_pf}

\subsection{Proof of Theorem \ref{gmnl_fptas}}\label{gmnl_fptas_pf}

Let $S^*$ be the optimal solution to the assortment optimization problem. There exists $l$ such that
$$
v(1 + \epsilon)^{l-1} \leq \sum_{j \in S^*} v_j \leq v(1+\epsilon)^{l}.
$$
Let $h = v(1+\epsilon)^l$. Then 
$$
\sum_{j \in S^*} \frac{v_j}{\epsilon h / n} \le \frac{n}{\epsilon h} h = \frac{n}{\epsilon},
$$
and rounding up gives us
$$
\sum_{j \in S^*}  \bar{v}_j \le \left\lceil \frac{n}{\epsilon} \right\rceil +n = I.
$$
Thus $S^*$ belongs to the set of assortments such that inequality $(1)$ is verified for $I$. Let $S_h$ be the assortment corresponding to $R(I,n)$ for the guess $h$, that is the one that maximizes $\sum_{j \in S} v_jp_j$ subject to $(1)$. Then since $S^*$ satisfies $(1)$ 
$$
\sum_{j \in S_h} v_jp_j \ge \sum_{j \in S^*} v_jp_j.
$$
Moreover,
$$
 \sum_{j \in S_h} v_j \leq \epsilon h /n \sum_{j \in S_h}  \bar v_j  \leq h(1+\epsilon + \epsilon/n) \le h(1+2\epsilon).
$$
Since $x \mapsto \frac{1}{x + v_0e^{\alpha (v_0 + x)}}$ is a decreasing function and
\begin{align*}
R(S_h) = \sum_{j \in S_h} \frac{v_j p_j}{\sum_{k \in S_h} v_k + v_0 e^{\alpha (v_0 + \sum_{j \in S_h} v_j)}} & \geq \frac{\sum_{j \in S_h} v_j p_j} {v(1+\epsilon)^{l}(1 + 2 \epsilon) + v_0 e^{\alpha (v_0 +v(1+\epsilon)^{l}(1 + 2 \epsilon))}}.
\end{align*}
Let us first show that there exists $\beta > 0$ such that
\begin{equation}
v(1+\epsilon)^{l-1} + v_0 e^{\alpha (v_0 + v(1+\epsilon)^{l-1}) }\geq (v(1+\epsilon)^{l}(1 + 2 \epsilon) + v_0 e^{\alpha (v_0 + v(1+\epsilon)^{l}(1 + 2 \epsilon))})\times (1 - \beta \epsilon).
\end{equation}
Indeed, 
\begin{align*}
   v(1+\epsilon)^{l-1} &\geq v(1+\epsilon)^{l}(1 + 2 \epsilon)(1 - \beta \epsilon) \\
  \Leftrightarrow  1 &\geq (1 + \epsilon)(1 + 2 \epsilon)(1-\beta \epsilon) = 1 + (3 - \beta) \epsilon + (2 - 3 \beta) \epsilon^2 - \beta  \epsilon^3,
\end{align*}
which is clearly true at least for $\beta \geq 3$. Moreover, 
\begin{align*}
    v_0 e^{\alpha (v_0 +v(1+\epsilon)^{l-1})}  &\geq v_0 e^{\alpha (v_0 + v(1+\epsilon)^{l}(1 + 2 \epsilon))}\times (1 - \beta \epsilon) \\
    \Leftrightarrow  e^{\alpha v(1+\epsilon)^{l-1}(1 - (1+\epsilon)(1 + 2 \epsilon))}  &\geq 1 - \beta \epsilon.
\end{align*}
Note that $e^{\alpha v(1+\epsilon)^{l-1}(1 - (1+\epsilon)(1 + 2 \epsilon))} = 1 - 3 \alpha v \epsilon + o(\epsilon)$. Therefore if we take $\beta \ge \max(3, 3 \alpha v)$, then the inequality $(2)$ is verified. 
Consequently
\begin{align*}
\frac{1}{v(1+\epsilon)^{l}(1 + 2 \epsilon) + v_0 e^{\alpha (v_0 + v(1+\epsilon)^{l}(1 + 2 \epsilon))}} &\geq \frac{1 - \beta \epsilon}{v(1+\epsilon)^{l-1} + v_0 e^{\alpha (v_0 + v(1+\epsilon)^{l-1})}} \\
&\geq \frac{1 - \beta \epsilon}{\sum_{k \in S^*} v_k + v_0 e^{\alpha (v_0 + \sum_{k \in S^*} v_k)}}.
\end{align*}

by definition of $l$. Therefore,
\begin{align*}
   R(S_h) & \geq (1 - \beta \epsilon) \frac{\sum_{j \in S_h} v_j p_j} {\sum_{k \in S^*} v_k + v_0 / e^{\alpha (v_0 + \sum_{k \in S^*} v_k)}} \\
R(S_h) & \geq (1 - \beta \epsilon)R(S^*),
\end{align*}
where in the last inequality we used that $\sum_{j \in S_h} v_jp_j \ge \sum_{j \in S^*} v_jp_j$. This proves that our algorithm returns a $(1-O(\epsilon))$-optimal solution to the assortment problem.

\noindent
\textbf{Running time} We try a total of $L = O(\log(nV/v)/\epsilon)$ guesses $h$, and for each guess we formulate a dynamic programming with $O(n^2/\epsilon)$ steps. Consequently the running time of the algorithm is $O\left( \frac{n^2}{\epsilon^2}\log(nV/v)\right)$ which is polynomial in the input size $n$ and $\frac{1}{\epsilon}$.

\subsection{Proof of Lemma \ref{lemma_rho_UV}}\label{lemma_rho_UV_pf}

Let $||.||$ be the usual Euclidean norm, and $\langle . \rangle$ its associated scalar product. We have that
$$
\rho(UV(S)) = \max_{x, ||x|| = 1} \langle UV(S)x, x \rangle.
$$
Let $k \in [K]$,
$$
\langle UV(S)e_k, e_k \rangle = \sum_{l = 1}^n (1 - \mu^{LR}(l,S)) u_{lk} v_{lk} \le \frac{1}{n} \sum_{l = 1}^n (1 - \mu^{LR}(l,S)),
$$
$$
\le \frac{1}{n} \left( n - |S| + \sum_{l \in S} (1 - e^{- \alpha \sum_{k \in [K]} u_{lk} V_k(S)}) \right),
$$
where the first inequality follows from the assumption made above on the coefficients $u_{lk} v_{lk}$. Furthermore, since $\alpha \le \log n$,
$$
e^{- \alpha \sum_{k \in [K]} u_{lk} V_k(S)} \ge e^{- \alpha \sum_{k \in [K]} u_{lk} \sum_{j \in N_+} v_{jk}} = e^{- \alpha} \ge \frac{1}{n}.
$$
Hence
$$
\langle UV(S)e_k, e_k \rangle \le \frac{1}{n} \left( n - |S| + |S| \left(1 - \frac{1}{n} \right) \right) = 1 - \frac{|S|}{n^2} \le 1 - \frac{1}{n^2}.
$$
The inequality holds for all $e_i, i \in [n]$, and therefore $\forall x \in \mathds{R}^K$ such that $||x|| = 1$. Hence,
$$
\rho(UV(S)) \le 1 - \frac{1}{n^2}.
$$

\subsection{Proof of Lemma \ref{gmnl2_exp_rev}}\label{gmnl2_exp_rev_pf}
Let $S \in N$ be the chosen subset of products, and let $i \in S$. We recall from Section \ref{choice_prob_form}, that
\begin{eqnarray}
\pi(i,S) = \lambda^T\left(I_n - Diag((1 - \mu(i,S))) \rho(N,N) \right)^{-1} \Pi(S) e_i.
\end{eqnarray}
This result still holds as the Generalized Multinomial Logit model with Low rank matrix has the same assumptions as the one needed to get this result. However, there is some difference with the assumptions made in Section \ref{sec_gmnl}. On the contrary of the Generalized Markov chain model presented in Section \ref{sec_gmnl}, we now have that
$$
\mu^{LR}(i,S) = e^{- \alpha \times \left( \sum_{k \in [K]} u_{ik} V_k(S) \right) }
$$
and
$$
\rho(N,N) = \sum_{k \in [K]} \mathbf{u}_k \mathbf{v}_k^T,
$$
where $\mathbf{v}_k = (v_{1k}, ..., v_{nk})$.
Therefore we have to compute the coefficients of the matrix 
$$\left(I_n - Diag((1 - \mu^{LR}(i,S))) \rho(N,N) \right)^{-1}$$
under this new model. We know that
\begin{eqnarray}
\left(I_n - Diag((1 - \mu^{LR}(i,S))) \rho(N,N) \right)^{-1} = \sum_{l = 0}^{\infty} \left( Diag((1 - \mu^{LR}(i,S))) \rho(N,N) \right)^{l}.
\end{eqnarray}
We use the following notations:
\begin{itemize}
\item let $M := Diag((1 - \mu^{LR}(i,S))) \rho(N,N) = \left( (1 - \mu^{LR}(i,S))\rho_{ij}\right)_{i,j \in [n]}$,
\item let $UV(S) = \left( \sum_{l = 1}^{n} (1 - \mu^{LR}(l,S)) u_{lk} v_{lm} \right)_{k,m \in [K]}
$,
\item for $l \in \mathds{N}$, we note $\left(UV(S)^{l}\right)_{k' k}$ the coefficient of indices $k',k$ of the matrix $UV(S)^{l}$, ie of the matrix $UV(S)$ elevated to the power of $l$.
\end{itemize} 
 
First, we show by induction that 
$$
\forall l \ge 1 \hspace{0.5cm} M^l = \left( (1 - \mu^{LR}(i,S)) \sum_{k,k' \in [K]} u_{ik} v_{jk'} \left(UV(S)^{l-1}\right)_{k' k} \right)_{i \in [n], j \in N_+}
$$
\textbf{Initiation} For $l = 1$, we use the definition of $M$ and $\rho(N,N)$ 
$$M = \left( (1 - \mu^{LR}(i,S))\rho_{ij}\right)_{i,j \in [n]} = \left( (1 - \mu^{LR}(i,S))\sum_{k \in [K]} u_{ik} v_{jk}\right)_{i,j \in [n]}.$$
Since $UV(S)^{0} = I_K$, we have that $\left(UV(S)^{0}\right)_{k' k} = \mathds{1}_{k' = k}$. Therefore
$$M = \left( (1 - \mu^{LR}(i,S))\sum_{k,k' \in [K]} u_{ik} v_{jk'}\left(UV(S)^{0}\right)_{k' k}\right)_{i,j \in [n]}.
$$
The result holds for $l = 1$.

\noindent
\textbf{Inductive step} Suppose that the result holds for a certain $l \in \mathds{N}^{*}$. We compute the coefficient of indices $i,j$ of $M^{l+1}$:
$$
\left( M^{l+1} \right)_{ij} = \sum_{q = 1}^n \left(M^l \right)_{i q} (1 - \mu^{LR}(q,S))\rho_{qj}
$$
Using the induction hypothesis, 
\begin{align*}
\left( M^{l+1} \right)_{ij} &=\sum_{q = 1}^n \left( (1 - \mu^{LR}(i,S)) \sum_{k,k' \in [K]} u_{ik} v_{qk'} \left(UV(S)^{l-1}\right)_{k' k}\right) (1 - \mu^{LR}(q,S)) \sum_{k'' \in [K]} u_{qk''} v_{jk''} \\
&= (1 - \mu^{LR}(i,S)) \sum_{k,k'' \in [K]} u_{ik} v_{jk''}  \sum_{k' \in [K]}  \left(\sum_{q \in [n]}(1 - \mu^{LR}(q,S)) u_{qk''} v_{qk'} \right) \left(UV(S)^{l-1}\right)_{k' k} \\
&= (1 - \mu^{LR}(i,S)) \sum_{k,k'' \in [K]} u_{ik} v_{jk''}  \sum_{k' \in [K]} (UV(S))_{k'' k'} \left(UV(S)^{l-1}\right)_{k' k} \\
&= (1 - \mu^{LR}(i,S)) \sum_{k,k'' \in [K]} u_{ik} v_{jk''} \left(UV(S)^{l}\right)_{k'' k}.
\end{align*}
Therefore the result holds for $l+1$.

\noindent
\textbf{Conclusion} For all $l \ge 1$, we have that
$$
M^l = \left( (1 - \mu^{LR}(i,S)) \sum_{k,k' \in [K]} u_{ik} v_{jk'} \left(UV(S)^{l-1}\right)_{k' k} \right)_{i \in [n], j \in N_+}
$$
Now that once we have this result, we can compute the coefficient of individual indices $i,j$ of the matrix $(I_n - Diag((1-\mu^{LR}(i,S)))\rho(N,N))^{-1} = (I_n - M)^{-1}$ using $(2)$:
\begin{align*}
\left((I_n - Diag((1-\mu^{LR}(i,S)))\rho(N,N))^{-1}\right)_{ij} &= \sum_{l = 0}^{\infty} (M^l)_{ij} \\
&= \mathds{1}_{i = j} + \sum_{l = 1}^{\infty} (1 - \mu^{LR}(i,S)) \sum_{k,k' \in [K]} u_{ik} v_{jk'} \left(UV(S)^{l-1}\right)_{k' k} \\
&= \mathds{1}_{i = j} + (1 - \mu^{LR}(i,S)) \sum_{k,k' \in [K]} u_{ik} v_{jk'} \sum_{l = 1}^{\infty} \left(UV(S)^{l-1}\right)_{k' k} \\
&= \mathds{1}_{i = j} + (1 - \mu^{LR}(i,S)) \sum_{k,k' \in [K]} u_{ik} v_{jk'} \left((I_K - UV(S))^{-1} \right)_{k'k}
\end{align*}.
The last inequality holds since $\rho(UV(S)) < 1$, as shown in Lemma~\ref{lemma_rho_UV}. Injecting it in $(1)$, and computing $\sum_{i \in S} \pi^{LR}(i,S)p_i$, we have
\begin{align*}
R^{LR}(S) &= \sum_{i = 1}^n \lambda_i  \sum_{j = 1}^n \left(\mathds{1}_{i = j} + (1 - \mu^{LR}(i,S)) \sum_{k,k' \in [K]} u_{ik} v_{jk'} \left((I_K - UV(S))^{-1} \right)_{k'k}\right) \mu^{LR}(j,S)p_j \\
&= \sum_{i = 1}^n \lambda_i \mu^{LR}(i,S)p_i + \sum_{i = 1}^n \lambda_i (1 - \mu^{LR}(i,S)) \sum_{j = 1}^n  \left( \sum_{k,k' \in [K]} u_{ik} v_{jk'} \left((I_K - UV(S))^{-1} \right)_{k'k}\right) \mu^{LR}(j,S)p_j \\
&= \sum_{i = 1}^n \lambda_i \mu^{LR}(i,S)p_i + \sum_{i = 1}^n \lambda_i (1 - \mu^{LR}(i,S)) \sum_{k,k' \in [K]} u_{ik} \left((I_K - UV(S))^{-1} \right)_{k'k} \sum_{j = 1}^n v_{jk'}\mu^{LR}(j,S)p_j \\
&= \sum_{i = 1}^n \lambda_i \sum_{k,k' \in [K]} u_{ik} \left((I_K - UV(S))^{-1} \right)_{k'k} \left( \sum_{j = 1}^n v_{jk'}\mu^{LR}(j,S)p_j \right) \\
&+ \sum_{i \in S} \lambda_i \mu^{LR}(i,S) \left(p_i -  \sum_{k,k' \in [K]} u_{ik} \left((I_K - UV(S))^{-1} \right)_{k'k} \left( \sum_{j = 1}^n v_{jk'}\mu^{LR}(j,S)p_j\right) \right).
\end{align*}
Since $\mu^{LR}(j,S)=0, \forall j\notin S$, after reorganizing the terms, we can rewrite this as
\begin{align*}
R^{LR}(S) = &\sum_{i \in [n]} \lambda_i (1 - \mu^{LR} (i,S)) \sum_{k,k' = 1}^{K} u_{ik} (I - UV(S))^{-1}_{k'k}\left(\sum_{l \in S} v_{lk'} \mu^{LR}(l,S) p_l\right) \\
&  + \sum_{i \in S} \lambda_i \mu^{LR}(i,S) p_i.
\end{align*}
Finally, writing the sum over $k,k'$ in matrix form and using the notation $\bu_i = \{u_{ik}\}_{k=1}^K$ and $\bv_j = \{v_{jk}\}_{k=1}^K$, we get
$$
R^{LR}(S) = \sum_{i \in [n]} \lambda_i (1 - \mu^{LR} (i,S)) \left(\sum_{j \in S} p_j \mu^{LR}(j,S) \bu_i^T[I - UV(S)]^{-1} \bv_{j} \right)   + \sum_{i \in S} \lambda_i \mu^{LR}(i,S) p_i,
$$
which concludes the proof.

\subsection{Proof of Lemma \ref{closeToUV}} \label{closeToUV_pf}
From Lemma~\ref{lemma_rho_UV}, we know that $\rho(UV(S)) \le 1 - \frac{1}{n^2} < 1$. Therefore we can write 
$$
[I - UV(S)]^{-1}  = \sum_{l = 0}^{\infty} \left(UV(S)^l\right). 
$$
We first show one side of the inequality. Let $\alpha_1 = (1-O(\epsilon))$. Furthermore, let $L = [I-UV(S)]^{-1} \bv_j$ and $L_\epsilon = [I-\alpha_1 UV(S)]^{-1} \alpha_1 \bv_j$. Hence, we have
\begin{align*}
    L-L_\epsilon & = \left([I-UV(S)]^{-1} - \alpha_1[I-\alpha_1 UV(S)]^{-1} \right)\bv_j \\
    & = \left(\sum_{l = 0}^{\infty} \left(UV(S)^l\right) - \alpha_1 \sum_{l = 0}^{\infty} \left(\alpha_1UV(S)^l\right)\right) \bv_j \\
    & = \left(  \sum_{l = 0}^{\infty} (1-\alpha_1^{l+1}) \left(UV(S)^l\right)\right) \bv_j \\
    & \geq \left(  \sum_{l = 0}^{\infty} (1-\alpha_1)^{l+1} \left(UV(S)^l\right)\right) \bv_j & \left(\text{since } 0 < \alpha_1 < 1 \right) \\
    & = (1-\alpha_1) \left(  \sum_{l = 0}^{\infty} (1-\alpha_1)^{l} \left(UV(S)^l\right)\right) \bv_j \\
    & = (1-\alpha_1) [I-(1-\alpha_1) UV(S)]^{-1} \bv_j \\
    & \geq O(\epsilon) L
\end{align*}
which completes the proof. The other side of the inequality is similarly proved using the fact that for $\alpha_1 >1$ (when we set $\alpha_1=(1+O(\epsilon)$), we have $\alpha_1^l - 1 \geq (\alpha_1-1)^l$ for any $l \geq 0$.

\subsection{Proof of Theorem \ref{fptas_gmnl2}} \label{fptas_gmnl2_pf}
Let $S^*$ be the optimal solution to the assortment optimization problem. There exist $t_1, ..., t_K$ such that for all $k \in [K]$
$$
v^k(1 + \epsilon)^{t_k} \leq \sum_{j \in S^*} v_{jk} =: V_k(S^*) \leq v^k(1+\epsilon)^{t_k + 1}.
$$
Let $h = (v^1(1+\epsilon)^{t_1}, ..., v^K(1 + \epsilon)^{t_K})$. Choose the set $S_h$ that maximizes the dynamic program defined above. Then by definition of $\bar v_{jk}$, for each $k \in [K]$
$$
 V_k(S_h) := \sum_{j \in S_h} v_{jk} \leq \epsilon h_k /n \sum_{j \in S_h}  \bar v_{jk}  \le \frac{\epsilon h_k }{n} U = \frac{\epsilon h_k }{n} \left(\lceil n/ \epsilon\rceil + n \right) \le  h_k(1+2\epsilon),
$$
and 
$$
V_k(S_h) := \sum_{j \in S_h} v_{jk} \ge \epsilon h_k / n \sum_{j \in S_h} (\bar v_{jk} - 1) \ge \frac{\epsilon h_k}{n}(L - |S_h|) \ge  \frac{\epsilon h_k}{n}(\lceil n / \epsilon \rceil  - n) \ge h_k (1 - \epsilon ).
$$
First of all, since for all $k \in [K]$, $h_k \le V_k(S^*) \le h_k(1+ \epsilon)$,
$$
e^{- \alpha \sum_{k = 1}^K h_k u_{jk} (1 + \epsilon)} \le \mu^{LR}(j,S^*) = e^{- \alpha \sum_{k = 1}^K V_k(S^*)u_{jk}} \le e^{- \alpha \sum_{k = 1}^K h_k u_{jk}} = \mu_j(h).
$$
Thus, if we note $\sum_{k = 1}^K h_k u_{jk} =: \langle h, u_j \rangle$, 
\begin{equation}\label{mu_bd}
\mu_j(h)(1 - \epsilon \alpha \langle h, u_j \rangle) \le \mu^{LR}(j, S^*) \le \mu_j(h),
\end{equation}
and using the same arguments for $h_k(1 - \epsilon) \le V_k(S_h) \le h_k(1+2\epsilon)$, we can show that 
\begin{equation*}
    \mu_{j}(h)(1 - 2\epsilon \alpha \langle h, u_j \rangle) \le \mu^{LR}(j,S_h) \le \mu_{j}(h) (1 + \epsilon \alpha \langle h, u_j \rangle).
\end{equation*}
Therefore
\begin{align*}
UV(S^*)_{kk'} = \sum_{i \in [n]} (1 - \mu^{LR}(i,S^*))u_{ik} v_{ik'} &\le \sum_{i \in [n]} (1 - \mu_i(h))u_{ik} v_{ik'}\left(1 + \frac{\mu_i(h)}{1 - \mu_i(h)} \epsilon \alpha \langle h, u_i \rangle \right) \\
&\le \sum_{i \in [n]} (1 - \mu_i(h))u_{ik} v_{ik'}\left(1 + \epsilon \alpha \max_{i \in [n]} \left\{ \frac{\mu_i(h)}{1 - \mu_i(h)}  \langle h, u_i \rangle \right\} \right)  \\
\end{align*}
Hence we get for all $k,k' \in [K]$
\begin{equation}\label{uv_bd}
    H(h)_{kk'} \le UV(S^*)_{kk'} \le H(h)_{kk'} (1 + \delta),
\end{equation}
where $\delta := \epsilon \alpha \max_{i \in [n]} \left\{ \frac{\mu_i(h)}{1 - \mu_i(h)}  \langle h, u_i \rangle \right\} $, and likewise 
\begin{equation*}
     H(h)_{kk'}(1 - \delta) \le UV(S_h)_{kk'} \le H(h)_{kk'} (1 + 2\delta).
\end{equation*}
Since $\delta = O(\epsilon)$, we can use the result from Lemma \ref{closeToUV}.
% this implies that  for all $k,k' \in [K]$
% \begin{equation}\label{uv_bd2}
% \left((I - H(h))^{-1}\right)_{{kk'}} \le \left((I - UV(S^*))^{-1}\right)_{{kk'}} \le \left((I - H(h))^{-1}\right)_{{kk'}}(1 + \delta X(H)),
% \end{equation}
% and 
% \begin{equation*}
%     \left((I - H(h))^{-1}\right)_{{kk'}}(1 - \delta X(h)) \le \left((I - UV(S_h))^{-1}\right)_{{kk'}} \le \left((I - H(h))^{-1}\right)_{{kk'}}(1 + 2\delta X(H)).
% \end{equation*}
We want to prove that $R^{LR}(S_h) \ge (1-O(\epsilon))R^{LR}(S^*)$.
% (1-g(\epsilon))R^{LR}(S^*)$ for a certain function $g$ such that $g(x) \underset{x \rightarrow 0}{\rightarrow }0$. 
Recall the expression of optimal revenue
\begin{equation*}
   R^{LR}(S^*) = \sum_{i\in S^*} \lambda_i \mu^{LR}(i,S^*) p_i + \sum_{i=1}^n \lambda_i (1-\mu^{LR}(i,S^*))f(i,S^*).
\end{equation*}
Let us now compare $f_i(h,S^*)$ and $f(i,S^*) := \sum_{j = 1}^n p_j \mu^{LR}(j,S^*) \bu_i^T [I-UV(S^*)]^{-1} \bv_j $
% \sum_{k,k' = 1}^K u_{ik}(I-UV(S^*))^{-1}_{k'k} \left( \sum_{j = 1}^n v_{jk'} \mu^{LR}(j,S^*)p_j \right)$ 
using bounds from (\ref{mu_bd}) and the result from Lemma \eqref{closeToUV}. We have
$$
f_i(h,S^*) (1 - \delta_2) \le f(i,S^*) := \sum_{j = 1}^n p_j \mu^{LR}(j,S^*) \bu_i^T [I-UV(S^*)]^{-1} \bv_j
% \sum_{k,k' = 1}^K u_{ik}(I-UV(S^*))^{-1}_{k'k} \left( \sum_{j = 1}^n v_{jk'} \mu^{LR}(j,S^*)p_j \right)
\le f_i(h,S^*) (1 + O(\epsilon)),
$$
where $\delta_2 := \epsilon \alpha \max_{i \in [n]} \left\{ \langle h, u_i \rangle\right\} = O(\epsilon)$. Using the bounds for $\mu^{LR}(j,S_h)$ and $f(i,S_h)$, we also have
\begin{equation*}
    f_i(h,S_h) (1 - O(\epsilon))(1 - 2 \delta_2) \le f(i,S_h) \le f_i(h,S_h) (1 + O(\epsilon)(1 + \delta_2).
\end{equation*}
Using these upper and lower bounds for $f(i,S^*)$ (resp. $f(i,S_h)$) and the corresponding lower and upper bounds for $\mu^{LR}(j, S^*)$ (resp. $\mu^{LR}(j,S_h)$) from equation (\ref{mu_bd}), we have 
\begin{align*}
    R^{LR}(S^*) & \le \sum_{i\in S^*} \lambda_i \mu_i(h) p_i + \sum_{i=1}^n \lambda_i (1-\mu_i(h)) (1+\delta)f_i(h,S^*) \left(1 + O(\epsilon)\right) \\
    & \le (1+\delta)\left(1 + O(\epsilon)\right) R^{DP}(h,S^*),
\end{align*}
and
\begin{align*}
    R^{LR}(S_h) & \ge \sum_{i\in S^*} \lambda_i (1 - 2 \delta_2) \mu_i(h) p_i + \sum_{i=1}^n \lambda_i (1-\mu_i(h))(1 - \delta)(1 - O(\epsilon)) (1-2\delta_2)f_i(h,S_h) \\
    & \ge (1 - \delta)(1 - O(\epsilon))(1-2\delta_2) R^{DP}(h,S_h).
\end{align*}
Next, by the definition of $S_h$ being the assortment that maximizes the dynamic program for the given guess $h$, we also have $R^{DP}(h,S^*)\leq R^{DP}(h,S_h)$.
Combining this with the bounds derived above, we get
\begin{equation}\label{upper_bd1}
    R^{LR}(S^*) \leq \frac{(1+\delta)(1+O(\epsilon))}{(1 - \delta)(1 - O(\epsilon))(1-2\delta_2)} R^{LR}(S_h).
\end{equation}
Hence we get the following lower bound for the expected revenue obtained by the assortment $S_h$:
\begin{equation*}
    \frac{(1 - \delta)(1 - O(\epsilon))(1-2\delta_2)}{(1+\delta)(1+O(\epsilon))} R^{LR}(S^*) \leq R^{LR}(S_h).
\end{equation*}
Finally, to argue that this gives us $(1-O(\epsilon))$ solution, 
% since $\delta X(H)  > 0$, 
to conclude the proof, we need to show that $\frac{(1 - \delta)(1 - O(\epsilon)) (1-2\delta_2)} {(1+\delta)(1+O(\epsilon))}$ is not too far from 1. This holds since,

$$
\delta :=O(\epsilon)  \hspace{0.5cm} \mbox{ and } \hspace{0.5cm} \delta_2 :=O(\epsilon).
$$
Thus 
% $$\delta X(H) = C_3(h) \epsilon,$$and 
$$
\frac{(1 - \delta)(1 - O(\epsilon))(1-2\delta_2)}{(1+\delta)(1 +O(\epsilon))} 
% = 1 - 2(C_1(h)+C_2(h)+C_3(h)) \epsilon + o(\epsilon) 
\ge 1 - O(\epsilon),
$$
% where $\beta$ is a constant and $C_1(h), C_2(h), C_3(h)$ are polynomial functions of the $h_k$'s. The last inequality holds because $C_1(h), C_2(h)$ and $C_3(h)$ can be easily bounded over $W_\epsilon$. The bound may depend on $\epsilon$, but since we multiply it by $\epsilon$, it will again be an $o(\epsilon)$ term, and is thus of no importance. 

We have finally proven that our algorithm returns an assortment $S_h$ that is a $(1-O(\epsilon))$ optimal solution to our assortment problem.
\begin{equation}\label{rev_bounds}
    R^{LR}(S^*)(1 - O(\epsilon)) \le R^{LR}(S_h) \le R^{LR}(S^*)
\end{equation}

\noindent
\textbf{Running time} We try a total of $\prod_{k \in [K]} O(\log(nV^k/v^k)/\epsilon) = O((\log(nV/v)/\epsilon)^K)$ guesses for $h$ (where $V := \max_{k \in [K]} V^k$ and $v := \min_{k \in [K]} v^k$). For each guess we formulate a dynamic programming with $O\left(\left(\frac{n}{\epsilon}\right)^{2K}n^2 K^2\right)$ run-time. Consequently the running time of the algorithm is $O\left(\frac{n^{2(K+1)} K^2}{\epsilon^{3K}}\log^K(nV/v)\right)$ which is polynomial in the input size $n$ and $\frac{1}{\epsilon}$.

\section{Bibliography}

\begin{spacing}{1}
{\small
\bibliographystyle{plain}
\bibliography{references_list}
}
\end{spacing}
 
\end{document}